\documentclass{iacrtrans}
\usepackage[utf8]{inputenc}
\usepackage{xspace}
\usepackage{listings}
\usepackage{pifont}
\usepackage{tugraz_defaults}
\usepackage{makecell}
\usepackage{subfig}

\usepackage{xcolor}
\definecolor{vgreen}{RGB}{104,180,104}
\definecolor{vblue}{RGB}{49,49,255}
\definecolor{vorange}{RGB}{255,143,102}

\lstdefinestyle{verilog-style}
{
    language=Verilog,
    basicstyle=\small\ttfamily,
    keywordstyle=\color{vblue},
    identifierstyle=\color{black},
    commentstyle=\color{vgreen},
    morekeywords = [2]{
      typedef,
      enum
     }
}

\colorlet{punct}{red!60!black}
\definecolor{background}{HTML}{EEEEEE}
\definecolor{delim}{RGB}{20,105,176}
\colorlet{numb}{magenta!60!black}

\lstdefinelanguage{json}{
    basicstyle=\normalfont\ttfamily,
    numbers=left,
    numberstyle=\scriptsize,
    stepnumber=1,
    numbersep=8pt,
    showstringspaces=false,
    breaklines=true,
    frame=lines,
    literate=
     *{0}{{{\color{numb}0}}}{1}
      {1}{{{\color{numb}1}}}{1}
      {2}{{{\color{numb}2}}}{1}
      {3}{{{\color{numb}3}}}{1}
      {4}{{{\color{numb}4}}}{1}
      {5}{{{\color{numb}5}}}{1}
      {6}{{{\color{numb}6}}}{1}
      {7}{{{\color{numb}7}}}{1}
      {8}{{{\color{numb}8}}}{1}
      {9}{{{\color{numb}9}}}{1}
      {:}{{{\color{punct}{:}}}}{1}
      {,}{{{\color{punct}{,}}}}{1}
      {\{}{{{\color{delim}{\{}}}}{1}
      {\}}{{{\color{delim}{\}}}}}{1}
      {[}{{{\color{delim}{[}}}}{1}
      {]}{{{\color{delim}{]}}}}{1},
}

\makeatletter
\newcommand*\@lbracket{[}
\newcommand*\@rbracket{]}
\newcommand*\@colon{:}
\newcommand*\colorIndex{%
    \edef\@temp{\the\lst@token}%
    \ifx\@temp\@lbracket \color{black}%
    \else\ifx\@temp\@rbracket \color{black}%
    \else\ifx\@temp\@colon \color{black}%
    \else \color{vorange}%
    \fi\fi\fi
}
\makeatother

\newif\ifanonymous

\newcommand\otfi{SYNFI\xspace}
\newcommand\ot{OpenTitan\xspace}
\ifanonymous
\newcommand\link{Repository link hidden for the blind review process.}
\else
\newcommand\link{https://github.com/lowRISC/synfi}
\fi

\definecolor{red_otfi}{HTML}{DB4437}
\definecolor{blue_otfi}{HTML}{4285F4}
\definecolor{green_otfi}{HTML}{0F9D58}

\ifanonymous
\author{}
\else
\author{Pascal Nasahl\footnote{The work was done while the author was at Google.}\inst{1,3} \and Miguel Osorio\inst{1} \and Pirmin Vogel\inst{2} \and Michael Schaffner\inst{1} \and Timothy Trippel\inst{1} \and Dominic Rizzo\inst{1} \and Stefan Mangard\inst{3,4}}
\institute{Google, Mountain View, USA \and lowRISC CIC, Cambridge, United Kingdom \and Graz University of Technology, Graz, Austria \\\href{mailto:pascal.nasahl@iaik.tugraz.at,stefan.mangard@iaik.tugraz.at}{firstname.lastname@iaik.tugraz.at} \and Lamarr Security Research, Graz, Austria}

\authorrunning{Nasahl et al.}
\fi
\title{SYNFI: Pre-Silicon Fault Analysis of an Open-Source Secure Element}

\begin{document}

\maketitle


\begin{abstract}
Fault attacks are active, physical attacks that an adversary can leverage to alter the control-flow of embedded devices to gain access to sensitive information or bypass protection mechanisms.
Due to the severity of these attacks, manufacturers deploy hardware-based fault defenses into security-critical systems, such as secure elements.
The development of these countermeasures is a challenging task due to the complex interplay of circuit components and because contemporary design automation tools tend to optimize inserted structures away, thereby defeating their purpose.
Hence, it is critical that such countermeasures are rigorously verified \emph{post-synthesis}.
Since classical functional verification techniques fall short of assessing the effectiveness of countermeasures (due to the circuit being analyzed when no faults are present), developers have to resort to methods capable of injecting faults in a simulation testbench or into a physical chip sample.
However, developing test sequences to inject faults in simulation is an error-prone task and performing fault attacks on a chip requires specialized equipment and is incredibly time-consuming.
Moreover, identifying the fault-vulnerable circuit is hard in both approaches, and fixing potential design flaws post-silicon is usually infeasible since that would require another tape-out.
%
%
To that end, this paper introduces \otfi, a formal pre-silicon fault verification framework that operates on \emph{synthesized} netlists.
\otfi can be used to analyze the general effect of faults on the input-output relationship in a circuit and its fault countermeasures, and thus enables hardware designers to assess and verify the effectiveness of embedded countermeasures in a systematic and semi-automatic way.
The framework automatically extracts sensitive parts of the circuit, induces faults into the extracted subcircuit, and analyzes the faults' effects using formal methods.
To demonstrate that \otfi is capable of handling unmodified, industry-grade netlists synthesized with commercial and open tools, we analyze \ot, the first open-source secure element.
In our analysis, we identified critical security weaknesses in the unprotected AES block, developed targeted countermeasures, reassessed their security, and contributed these countermeasures back to the \ot project.
For other fault-hardened IP, such as the life cycle controller, we used \otfi to confirm that existing countermeasures provide adequate protection.
%

\end{abstract}

\section{Introduction}
\label{sec:otfi:introduction}

In a fault attack, an adversary induces a fault into a chip to manipulate the execution of the circuit.
The physical effect of the fault can then be exploited to hijack the control-flow of a CPU~\cite{seaborn2015exploiting, nasahl2019attacking}, to bypass secure-boot~\cite{vasselle2017laser}, or to leak sensitive data of the device~\cite{biham1997differential, piret2003differential, dobraunig2018sifa}.
Fault attackers often target secure elements, as they handle highly security-sensitive assets.
For this reason, these root-of-trust~(RoT) elements, such as \ot~\cite{johnson2018titan}, embed several hardware-based fault hardening techniques into the chip.
As the resistance of the circuit against faults relies on these countermeasures, it must be assured that they provide the expected security guarantees.

To ensure the correctness of the countermeasures, hardware engineers responsible for designing secure chips must analyze the circuit when influenced by faults in the design phase.
This pre-silicon evaluation needs to comprise two central analysis points:
First, \textit{can induced faults influence the input-output relation of a security-critical circuit and can the countermeasures detect them?}
Here, the hardware designer wants to reveal whether a fault affects the circuit and to verify that the countermeasure achieves the promised security level, \ie can handle up to a certain number of simultaneously induced faults specified in the threat model.
Second, \textit{can the embedded countermeasures hinder an adversary from entering a specific, security-critical circuit state using faults?}
An example of such a state is the debug mode of a secure element allowing the adversary to escalate privileges.

Testing the resilience of the circuit and its countermeasures against faults needs to be conducted in the last stage of the front-end design, \ie at synthesized gate-level netlist.
This approach ensures that \textit{(i)} defective countermeasures are detected as early as possible avoiding long design turnaround times.
Additionally, at this \textit{(ii)} level of abstraction, the design uses the standard cell library provided by the manufacturer and, therefore, is already close to the final circuit sent to the fab for the tape-out.
Furthermore, performing the security assessment at the netlist ensures \textit{(iii)} that flaws introduced by the tooling can be detected.
Here, especially the logic synthesis design flow step mapping the register-transfer level~(RTL) model to the synthesized gate-level netlist could negatively affect countermeasures using redundancy to detect or mitigate faults.
Here, the synthesis optimization passes aiming to meet design constraints, \eg the area consumption, could be responsible of reducing security guarantees.

One approach of analyzing the resilience of the circuit against faults at the netlist level is to manually induce faults and to analyze their effect in the simulation phase.
However, as the names of the wires and cells in the netlist are renamed or mangled by synthesis tools, manually inducing faults in the testbench is an error-prone task.
Additionally, the analysis process is very time-consuming since the simulation needs to be restarted for each induced fault. 
Hence, this process is often at risk of being foregone in the verification phase of the design due to development schedule pressure.

In order to verify the functionality of fault countermeasures embedded into the chip, a framework capable of automatically performing a pre-silicon analysis based on the synthesized gate-level netlist is needed.
It is crucial for such a tool to be capable of handling industry-grade netlists using proprietary standard-cell libraries without imposing any restrictions on the netlist and the design.
Otherwise, such restrictions would render the tool practically irrelevant, especially for commercial projects that rely on established hardware design flows with a multi-stakeholder design team.

Recently published tools~\cite{burchard2017autofault, arribas2020cryptographic, richter2021fiver, bosio2008lifting, simevski2013automated} cannot be used to analyze unmodified netlists of industry-driven projects, as these frameworks impose invasive requirements to the design.
Tools, such as FIVER~\cite{richter2021fiver}, limit \textit{(a)}, the supported gates in the netlist to a small set, preventing the usage of complex, proprietary standard cell libraries.
Furthermore, most of these frameworks~\cite{richter2021fiver} require \textit{(b)} that the given netlist does not include any cycles, \ie the hardware designer needs to manually unroll the design before the evaluation.
Additionally, related work often demands \textit{(c)} that the netlist is fully flattened, \ie does not include any submodules or hierarchy, does not support \textit{(d)} all language features, or is not \textit{(e)} open-source.  
Finally, as most fault injection frameworks~\cite{burchard2017autofault, arribas2020cryptographic, bosio2008lifting} exclusively focus on analyzing cryptographic primitives \textit{(f)}, it remains unclear whether these frameworks also can be used to assess the security of more generic hardware designs consisting of a diverse set of hardware IP components, especially in respect with the two analysis points described above.

\subsection*{Contribution}
In this paper, we present \otfi, a versatile framework capable of performing a pre-silicon fault analysis of synthesized gate-level netlists.
\otfi allows a hardware designer to automatically analyze the resilience of a circuit and its countermeasures against fault attacks with minimal setup overhead.
More specifically, \otfi enables hardware designers and security engineers to study the impact of faults on the circuit, to analyze the functionality of tailored fault countermeasures, and to investigate which cells are the most critical attack targets and need special protection.
This information can be used to find logical flaws in the design as well as defects introduced by the hardware design flow tools before the tape-out of the chip.

The \otfi framework is capable of performing the pre-silicon fault analysis on unmodified netlists generated with proprietary or open design flows and standard cell libraries of designs using common hardware design patterns.
For the fault experiment, the security engineer needs to provide information about the circuit to analyze and the fault model.
\otfi supports fault models comprising single and multiple faults injected into various locations in the circuit and different fault effects, \ie transient or stuck-at effects.
With this configuration, \otfi automatically extracts the circuit to analyze from the netlist and injects faults according to the fault model.
In the analysis phase, \otfi reveals whether a fault affects the input-output relation of the circuit, shows whether the embedded countermeasures can detect faults up to a certain number, and verifies whether a fault could enable an adversary to enter a security-critical state.



To emphasize the importance of conducting a pre-silicon fault analysis before an upcoming tape-out, we utilize \otfi to analyze components of the \ot secure element.
In particular, we focus on analyzing the fault-resiliency of the most security-critical components, such as the AES primitive, the life cycle controller, the lockstep mode of the processor, and several other fault hardened IP.
For our assessment, we study the impact of single and multiple faults induced into different parts of the modules for various fault effects.
We conduct our analysis on the unmodified netlist generated with the internal, proprietary hardware design flow of \ot including a commercial standard cell library as well as on the netlist synthesized with open-source tools.
We utilize \otfi to \textit{(i)} reveal the impact of faults to unprotected circuits, to \textit{(ii)} verify that the redundancy-based countermeasures are not removed by the synthesis tool, and \textit{(iii)} to verify whether certain security-critical states cannot be entered using faults without triggering the countermeasures.
Our in-depth analysis of the tested modules revealed that the AES module is highly susceptible to fault attacks.
More concretely, our evaluation disclosed that already a single fault into the AES round counter, the handshake signals, or certain finite-state machines allow an adversary to break the security of the module.
To mitigate the encountered security violations, we developed several fault hardening mechanism and integrated them into the \ot project.
We ensured the correctness of these countermeasures by reassessing the hardened module using \otfi.
For fault-hardened modules, such as the life cycle controller, we were able to formally verify the expected fault-resiliency.


In summary, our contributions are:
\begin{itemize}
    \item We present and implement \otfi, an open-source\footnote{\link} framework capable of performing a pre-silicon fault analysis at the gate-level. \otfi allows security engineers to automatically \textit{(i)} reveal whether a fault affects the input-output relation of a circuit and its countermeasures and \textit{(ii)} assess if an adversary can enter a particular circuit state without triggering the countermeasures. In contrast to related work, the \otfi framework is able to process unmodified netlists of hardware designs making use of a variety of design patterns and synthesized with commercial and open-source synthesis tools.
    \item We identified several fault attack vectors for the unprotected AES module used in the \ot secure element allowing an adversary to threaten the security of the encryption primitive. To prevent the exploitation of these flaws in the final taped-out chip, we implemented, reassessed, and contributed several fault hardening techniques to the upstream project.
    \item We verified with \otfi that a selection of the most security-critical \ot IP blocks hardened against faults provides the expected security. In particular, we verified, among other modules, that an adversary cannot hijack the life cycle controller to enter the RMA debug state from the production state and that a fault into the program counter of the processor is detected by the lockstep mode of the CPU.
\end{itemize}


\section{Background}
\label{sec:otfi:background}
In this section, we summarize fault attacks and provide detailed background on the \ot project.
\subsection{Fault Attacks}
\label{sec:otfi:background:fi}
Fault attacks are active, physical attacks that are commonly used to threaten the security of embedded devices~\cite{debusschere2012modern,timmers2017escalating, o2020bam, elmohr2020fault} and secure elements~\cite{heriveauxdefeating, van2011practical, schink2021security}.
In these attacks, a fault is induced into the chip causing several effects at the physical level, \eg transient voltage and current changes as well as timing violations~\cite{richter2021revisiting}.
These side-effects are then exploited allowing an adversary to bypass security measures~\cite{vasselle2017laser, timmers2017escalating}, attack cryptographic primitives~\cite{biham1997differential, piret2003differential, dobraunig2018sifa}, or redirect the control-flow~\cite{seaborn2015exploiting, nasahl2019attacking} of the executed software.
The fault model, which is used to characterize such attacks, comprises the fault methodology, the spatial and temporal properties, and the effect of the fault.
The spatial and temporal properties of the fault model define the location, the duration, and the time of the induced fault.
Although, depending on these different fault parameters, the effect of a fault varies, commonly bit-flips and stuck-at effects are observed~\cite{riesgo1996fault}.
To induce a glitch into a system, various fault methodologies, such as voltage, clock, laser, and EM glitching~\cite{karaklajic2013hardware}, emerged in the previous years. 
While these fault methodologies originally only could be performed locally, new fault methodologies even allow inducing faults in software over the network~\cite{murdock2020plundervolt, tang2017clkscrew, qiu2019voltjockey}.

\subsection{OpenTitan}
\label{sec:otfi:background:ot}
Secure elements and RoT chips are used in smartphones~\cite{li_2020}, computers~\cite{applet2, titanc}, and servers~\cite{aws} to establish a secure anchor point.
These elements are trusted by the system and offer various services, such as cryptographic functions, key storage and support for secure boot protocols.
As a security breach could be fatal, these integrated circuits typically offer a certain level of protection~\cite{roche2021side} against fault attacks.
RoT chips introduced so far are closed, proprietary designs making it necessary for the system integrator to trust the manufacturer of these devices.
The \ot~\cite{johnson2018titan} project aims to obviate this requirement by providing the first open-source root-of-trust chip.
However, as the silicon design of the chip is open-source, an attacker also could discover potential attack vectors.
Hence, it must be assured that the installed countermeasures work as intended by using a rigorous verification approach.


\section{Design and Implementation}
\label{sec:otfi:design}

This section describes the fundamental concepts of the \otfi framework along with the design rationale.
We first give a high-level overview of the framework and then provide an in-depth description of all the design stages of \otfi.

\subsection{Overview}

To analyze the effects of one or multiple faults to the input-output relation of a circuit and its fault countermeasures, the gate-level netlist, the used standard cell library, as well as a fault specification need to be provided to the \otfi framework.\newline
\textbf{Netlist \& Cell Library:}
The first input of the \otfi framework is the unmodified netlist of the module to analyze and the standard cell library that the design is mapped against.
\begin{figure*}[h]
  \centering
  \includegraphics[width=\textwidth]{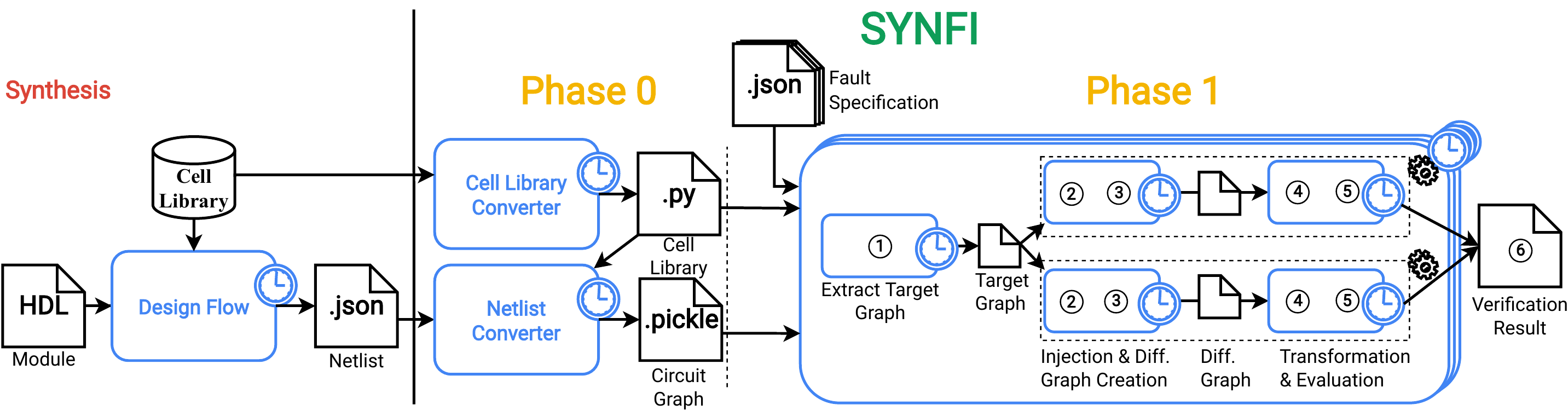}
  \caption{Block diagram of the \otfi framework.}
  \label{fig:otfi:blockdiag}
\end{figure*}
As shown in the block diagram in Figure~\ref{fig:otfi:blockdiag}, the synthesis design flow step, which is not part of the \otfi framework, is responsible for transforming the RTL design into the netlist using the standard cell library.\newline
\textbf{Fault Specification:}
The second input of the framework is the fault specification file responsible for describing the fault experiment the designer wants to perform.
\begin{lstlisting}[language=json, caption= {Fault specification.}, label={lst:otfi:faultmodel}]
"Fault Specification": {
    "Target Circuit": {
        "inputs": ["in_port1": "2'b00", "in_port2": "2'b01"],
        "outputs": ["out_port1": "2'b11", "in_port2": "2'b10"]
    },
    "Fault Model": {
        "Simultaneous Faults": 2,
        "Fault Locations": ["gate1", "gate2", "..."],
        "Fault Mappings": ["NAND2": ["AND2", "OR2"], "XNOR2": ["XOR2"]]
    }
}
\end{lstlisting}
As shown in Listing~\ref{lst:otfi:faultmodel}, the fault specification file is split into \textit{(i)} the description of the subcircuit the designer wants to evaluate and \textit{(ii)} the fault model describing the faults injected into this subcircuit.
The subcircuit to analyze \textit{(i)} is defined by the user by providing input and output nodes, \eg input or output ports, cells, or submodules of the design.
Furthermore, the user needs to assign inputs and expected output values for the provided nodes.
The fault model~\textit{(ii)} describes all faults which are induced into the subcircuit.
Here, the fault model consists of the \textit{(a)} number of faults injected into the circuit, the \textit{(b)} location, and the \textit{(c)} effects of the faults.
With the number of faults \textit{(a)}, the user can specify how many faults are simultaneously injected into the subcircuit.
In \otfi, a fault is injected into \textit{(b)} a certain location, \ie a gate.
Here, the user can either provide a list of gates which are attacked or select an exhaustive approach where \otfi automatically injects faults into all gates.
The last parameter is the \textit{(c)} fault effect.
Similar to~\cite{richter2021revisiting}, we model the effect of a fault induced into a certain gate by replacing the boolean function of the gate type according to a mapping.
For example, the mapping \texttt{NAND2=[AND2]} replaces a gate of type \texttt{NAND2} during the attack phase with an \texttt{AND2}.
Here, by inverting the boolean function, \otfi is capable of modeling a transient fault effect.
To model a stuck-at $0$ or $1$ fault, the boolean function of the corresponding gate can be set to a $0$ or $1$ in the fault mapping.
By providing multiple entries in the fault mapping, \eg \texttt{NAND2=[AND2, 0]}, \otfi can be used to analyze the circuit when influenced by transient or stuck-at faults.
Summarized, this mapping enables \otfi to model transient, stuck-at, or more advanced fault effects.
For each subcircuit the user wants to analyze, a new fault specification file needs to be provided and \otfi needs to be started again.
\newline
\textbf{SYNFI:}
With the unmodified netlist, the standard cell library, and the fault specification, the tool starts the two-phase transformation and analysis process depicted in Figure~\ref{fig:otfi:blockdiag}.
In \textit{Phase 0}, the framework transforms the netlist into a directed multigraph and converts the cell library into a format the subsequent steps of the \otfi framework support.
In \textit{Phase 1}, the subcircuit to analyze, \ie the target graph, is \ding{172} extracted from the circuit according to the fault specification file.
Afterwards, for each fault location and fault mapping combination, a separate process is started.
In these processes, two copies of the target graph are created, \ie the faulty and non-faulty target graph.
\otfi induces \ding{173} faults according to the fault mapping (number of simultaneous faults, location, and mapping) into the faulty target graph by replacing the boolean functions of the target gates according to the mapping.
By combining the faulty and non-faulty target and adding an input and output layer responsible for analyzing the effects of the induced faults, the differential graph \ding{174} is created.
This differential graph is used by \otfi to evaluate if a fault is effective, \ie the fault manipulates the outputs of the faulty target graph and is not detected by the countermeasures.
Although the detection of faults by the countermeasures is implementation specific, these countermeasures typically raise an error signal, which \otfi uses to evaluate whether the fault was detected or not.
Finally, the differential graph is converted to a boolean formula \ding{175} and a SAT solver utilizes this mathematical model representing the circuit to reason \ding{176} about the effectiveness of the induced faults.
In the end, the framework provides a detailed report \ding{177} summarizing the outcome of the fault analysis.

\subsection{Phase 0 - Cell Library \& Netlist Converter}
The first step the framework conducts is the transformation of the \textit{(i)} standard cell library and the \textit{(ii)} gate-level netlist.
The goal of this transformation step is to support arbitrary netlists generated with different hardware design flows and standard cell libraries.

First~\textit{(i)}, \otfi converts the provided standard cell library from the liberty format to a Python library.
\begin{lstlisting}[language=json, caption= {Cell library entry for an \texttt{AOI21\_X2} cell.}, label={lst:otfi:celllib},float,floatplacement=H]
"Cell Library": {
    "AOI21_X2": {
        "input_pins": ["A1", "B1", "B2"],
        "output_pins": "ZN",
        "boolean_function": "ZN = !(A1 & (B1 | B2))
    }
}
\end{lstlisting}
For this conversion, \otfi opens the provided standard cell library and extracts the name, the boolean function, and the input and output pins of the cells, as shown in Listing~\ref{lst:otfi:celllib}.
\otfi supports all cells with a boolean function, including compound gates, such as \texttt{AOI} cells.
Cells that are used due to their electrical rather than for their logical behavior, \eg filler cells, are not handled by \otfi as they are not used in the gate-level netlist.

Afterwards~\textit{(ii)}, \otfi transforms the unmodified netlist into a directed multigraph using a Python library~\cite{hagberg2008exploring}.
\begin{lstlisting}[language=json, caption= {Graph representation of the netlist.}, label={lst:otfi:netlistgraph}]
"Nodes": {
    "U1": { "type": "NAND2" },
    "U2": { "type": "AOI21" }
},
"Edges": {
	"1": {
		"out": { "node": "U1", "port": "ZN" },
		"in": { "node": "U2", "port": "A1" }
	}
}
\end{lstlisting}
In this graph, nodes represent ports, cells, and submodules and each of these nodes consists of a name and a type.
The type, \eg a \texttt{NAND2} gate or a port, defines the behavior of the node and the corresponding boolean function is provided by the cell library.
Similar to gates and ports, submodules are also represented as nodes and the corresponding boolean function needs to be provided by the user.
These nodes are connected using edges, which store information about the input and output port.
Listing~\ref{lst:otfi:netlistgraph} shows an example graph where the output port \texttt{ZN} of the gate \texttt{U1} is connected with the input port \texttt{A1} of the gate \texttt{U2}.

\subsection{Phase 1 - Target Graph Extraction}
\label{sec:otfi:targetextraction}
The target graph extraction \ding{172} step consists of the \textit{(i)} extraction and \textit{(ii)} preprocessing phase.

\subsubsection{Extraction}

The goal of the target graph extraction \textit{(i)} is to simplify the subsequent analysis phase by extracting the subcircuit the user wants to analyze with \otfi from the overall circuit.
The definition of the target graph is provided in the fault specification file.
Here, the user needs to define input and output nodes and the corresponding input and expected output values in the fault specification file.
These nodes can be any cell, port, or submodule in the circuit.

With this information, the \otfi framework starts the automatic target graph extraction process.
Here, the tool finds all paths, consisting of combinational and sequential logic, between the defined inputs and outputs.
Due to this extraction step, some nodes, \eg gates, are missing one or multiple inputs as the corresponding connecting gates are not part of the extracted circuit.
For all of these missing inputs, \otfi introduces auxiliary input nodes.

\subsubsection{Preprocessing}

The goal of the preprocessing phase \textit{(ii)} is to remove any time-dependencies in the extracted target graph.
This is necessary as the graph is converted into a time-independent mathematical model, \ie a boolean equation, in the last step described in Section~\ref{sec:otfi:design:transformation}.
In \otfi, we automatically break time-dependencies by \textit{(a)} replacing registers used in pipeline stages with pass-through elements and \textit{(b)} by removing loops and replacing registers in iterative designs and state machines.
These pass-through elements are time-independent, \ie do not have a clock port, and map the input to the (negated) output.
\otfi automatically distinguishes between the two register types by checking whether the register is the start and end of a cycle, \ie a register used in an iterative design.
This preprocessing phase enables \otfi to handle circuits that were not manually unrolled by the hardware designer.
However, when aiming to analyze multiple loop iterations, \eg multiple rounds in an iterative AES implementation, \otfi needs to evaluate each round individually.

\subsubsection{Preprocessing and Extraction Example}

\begin{figure}%
    \centering
    \subfloat[Example circuit with assigned input (blue) and output (green) values.\label{fig:otfi:cycles_1}]{\includegraphics[width=0.44\linewidth]{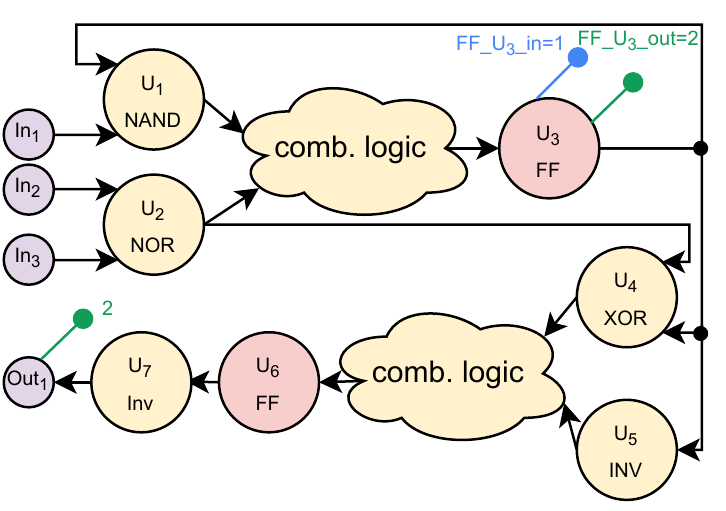}}%
    \qquad
    \subfloat[Extracted and preprocessed target graph.\label{fig:otfi:cycles_2}]{\includegraphics[width=0.49\linewidth]{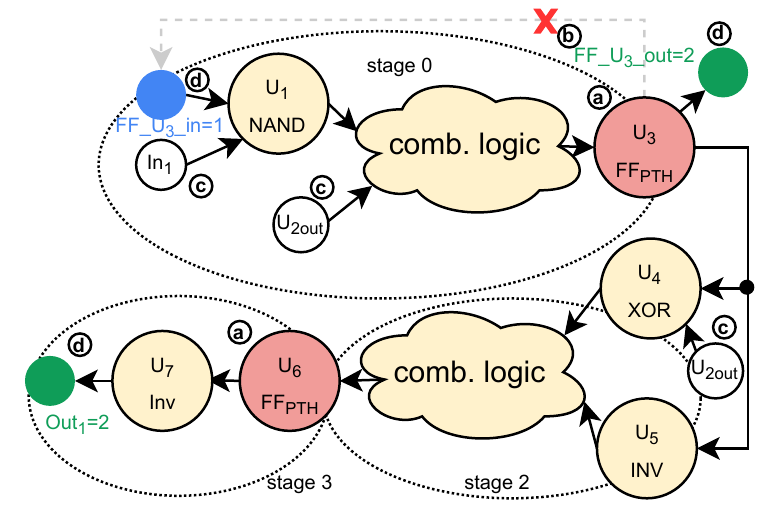}}%
    \caption{Target graph extraction.}%
    \label{fig:otfi:example}%
\end{figure}

We illustrate the extraction \textit{(i)} and preprocessing \textit{(ii)} phase in the example circuit in Figure~\ref{fig:otfi:example}.
Figure~\ref{fig:otfi:cycles_1} depicts a circuit consisting of three input ports ($In_1...In_3$), one output port ($Out_1$), two registers ($U_3, U_6$), and a set of combinational gates.
For the target graph extraction step \ding{172}, the user needs to provide input and output nodes and a corresponding circuit state, \ie values for these nodes.
In this example, we set the register $U_3=1$ (blue) as the input and the register $U_3=2$ and $Out_1=2$ (green) as the output in the fault specification file.

As the circuit contains a register used in a pipeline stage ($U_6$) and a register used in a sequential loop ($U_3$), \otfi removes these time-dependencies in the graph.
The registers are, as shown in Figure~\ref{fig:otfi:cycles_2}, replaced \textcircled{a} with pass-through elements and the loop between $U_3$ and $U_1$ is removed \textcircled{b}.
Then, \otfi finds all paths between the input node ($U_3$) and the output nodes ($U_3$, $Out_1$), \ie all nodes except $U_2, In_1, In_2$, and $In_3$.
To avoid that certain gates have unconnected inputs, \ie $U_1$, the frameworks adds auxiliary input nodes \textcircled{c} and connects them with the corresponding nodes. 
Finally, the framework adds input and output nodes \textcircled{d} for the user-defined input and output values.

\subsection{Phase 1 - Fault Injection}

After transforming the netlist into a graph and extracting the target graph, the injection \ding{173} phase starts.
For each fault combination, \ie number of simultaneous faults, fault locations, and fault mappings, defined in the fault model, \otfi starts a new process.
Inside these processes, \otfi creates two copies of the extracted target graph - the faulty and non-faulty target graph.
While the non-faulty target graph is used as a reference circuit in the subsequent steps, \otfi induces faults into the faulty target graph.
Here, the framework replaces the boolean function at a fault location according to the fault mapping.

To limit the configuration effort, \otfi already provides a default fault mapping for a large set of gates.
Furthermore, as the fault location is an optional parameter allowing the security engineer to attack specific gates, \otfi supports an exhaustive injection approach automatically targeting all available gates in the target graph.
Hence, at a minimum, the user only needs to specify the number of simultaneous faults injected into the gate-level netlist in the fault model.

\subsection{Phase 1 -  Differential Graph Creation}
\label{sec:otfi:diffgraph}

For each fault combination process, \otfi creates a differential graph \ding{174} consisting of a faulty and non-faulty target graph.
These differential graphs are responsible for evaluating the impact of faults on the circuit.
\begin{figure}[h]
  \centering
  \includegraphics[width=0.35\linewidth]{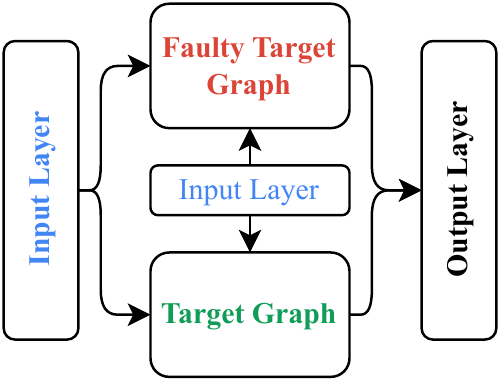}
  \caption{Differential graph.}
  \label{fig:otfi:diff_graph}
\end{figure}
As depicted in Figure~\ref{fig:otfi:diff_graph}, the differential graph consists of the faulty and non-faulty target graph.
To this differential graph, we add an input layer and an output layer.
In the input layer, we assign the input nodes added in the extraction phase \ding{172} the values provided by the user in the fault specification.
As the user does not need to provide all possible input values of the analyzed circuit, \otfi automatically connects the non-defined inputs of the faulty target graph with the non-defined inputs of the non-faulty target graph.
The SAT solver, which is used in the evaluation step and described in Section~\ref{sec:otfi:design:transformation}, then automatically assigns values to these non-defined inputs.
The output layer is used to analyze the effect of a fault.
This layer consists of a logic comparing the output values produced by the faulty and non-faulty target graph with the output values provided by the user in the fault specification.
Depending on the attack objectives and the implemented countermeasures, \otfi allows the hardware designer to define two different types of effective faults, which are defined by the impact of the fault to the output.
The detection of these two different effective fault types is implemented in the output layer.

\subsubsection{Unspecific Fault Effects}
\label{sec:otfi:arbitfault}
This type of effective fault enables \otfi to generically reveal whether a fault influences the input-output relation of a circuit.
For circuits without countermeasures, \otfi defines a fault to be \textbf{effective}, \textit{iff} this fault manipulates one or multiple output bits of the analyzed subcircuit, \ie the non-faulty and faulty target graph produce different output values.
\begin{equation}
\begin{split}
    OutputLogic = \color{green_otfi}(O_{NF} {=}{=} O_{E})\land \color{red_otfi}(O_{F} {!}{=} O_{E})
\end{split}
\label{eq:otfi:clause}
\end{equation}
Equation~\ref{eq:otfi:clause} depicts the logic in the output layer used to detect this type of fault effect.
The first part of the equation ensures that the output $O_{NF}$ of the non-faulty ($NF$) graph produces the expected output value $O_E$ defined in the fault specification.
More specifically, this part of the equation ensures that the SAT solver only assigns values to the non-defined inputs (\cf Section~\ref{sec:otfi:diffgraph}) of the differential graph which generate the expected output circuit state $O_E$.
The second part of the equation is responsible for ensuring that the output $O_{F}$ of the faulty ($F$) circuit does not match the expected output value.
If the output logic produces a logical $1$, an \textbf{effective} fault is found.
\begin{equation}
\begin{split}
    OutputLogic = \color{green_otfi}((O_{NF} {=}{=} O_{E}) \land (O_{NFA} {=}{=} 0)) \color{black} \land \color{red_otfi}((O_{F} {!}{=} O_{E}) \land (O_{FA} {=}{=} 0))
\end{split}
\label{eq:otfi:clause_alert}
\end{equation}
For circuits with dedicated fault countermeasures, \otfi considers a fault to be \textbf{effective}, \textit{iff} this fault manipulates one or multiple output bits of the analyzed subcircuit \textit{and} the alert signal of the countermeasure was not triggered.
If the alert signal was triggered, the countermeasure works as intended and the fault is considered to be \textbf{ineffective}.
The output logic in Equation~\ref{eq:otfi:clause_alert} models this behavior by ensuring that the alert signal $O_{FA}$ was not triggered in both parts of the formula. 

\subsubsection{Specific Fault Effects}
\label{sec:otfi:specfault}
This type of fault effect allows \otfi to check whether a fault enables the adversary to enter a specific circuit state.
Here, \otfi considers a fault to be \textbf{effective}, \textit{iff} the output of the faulty target graph matches the expected output value defined in the fault specification.
\begin{equation}
\begin{split}
    OutputLogic = \color{green_otfi}(O_{NF} {=}{=} O_{E}) \color{black} \land \color{red_otfi} (O_{F} {=}{=} O_{EF})
\end{split}
\label{eq:otfi:clause_expected fault}
\end{equation}
Equation~\ref{eq:otfi:clause_expected fault} shows the logic in the output layer capable of detecting this fault effect type.
Here, the first part of the equation ensures that the outputs $O_{NF}$ produced by the non-faulty target graph match the expected output values $O_{E}$ provided in the fault specification.
The second part of the equation ensures that the outputs of the faulty  graph match the expected fault output value $O_{EF}$ specified in the fault specification.
\begin{equation}
\begin{split}
    OutputLogic = \color{green_otfi}((O_{NF} {=}{=} O_{E}) \land (O_{NFA} {=}{=} O_{EA})) \color{black}\land \color{red_otfi}((O_{F} {=}{=} O_{EF}) \land (O_{FA} {=}{=} O_{EA}))
\end{split}
\label{eq:otfi:clause_expected fault_alert}
\end{equation}
For circuits consisting of a fault countermeasure designed to detect a fault, the alert signal $O_{EA}$ is also incorporated in the output logic, as shown in Equation~\ref{eq:otfi:clause_expected fault_alert}.
Here, the output layer ensures that the non-faulty circuit produces the expected output values $(O_{NF} {=}{=} O_{E_0})$ and that the alert was not triggered in the reference circuit, \ie $(O_{NFA} {=}{=} O_{EA})$.
For the faulty target graph, the equation ensures that the output value matches the expected fault output value $(O_{F} {=}{=} O_{EF})$, and the alert was not triggered.

\subsection{Phase 1 - Transformation \& Evaluation}
\label{sec:otfi:design:transformation}
After creating the differential graph, the \otfi framework converts this graph \ding{175} into a mathematical model.
As each node is assigned a boolean function, the tool uses the Tseitin transformation~\cite{tseitin1983complexity} to automatically transform the differential graph into a boolean formula in conjunctive normal form~(CNF).
The extracted boolean formula then is handed over to a SAT solver for the evaluation \ding{176}.
As shown in the differential graph in Figure~\ref{fig:otfi:diff_graph}, the inputs of the boolean formula are either set in the input layer to values provided by the user in the fault specification or are left unconnected.
For these unconnected input values, which are shared by the faulty and non-fault target graph, the SAT solver can set these values freely as long as the reference circuit produces the expected output values.
This is ensured by the output layer of the differential graph (\cf Section~\ref{sec:otfi:diffgraph}).
If the logic in the output layer produces a logical $1$, an effective fault is found.
For the report \ding{177}, the framework collects the number of effective faults, their location, and fault mapping.

\subsubsection{Selection of the SAT Solver}
For our Python-based tool, we use the PySAT~\cite{imms-sat18} framework as an interface to the SAT solver.
To determine the fastest solver for our purpose, we executed several fault injection verification experiments with the provided solvers~\cite{BiereFazekasFleuryHeisinger-SAT-Competition-2020-solvers, een2003extensible, Biere-SAT-Competition-2017-solvers, audemard2018glucose, minicard, liang2018maplesat} as a custom benchmark and decided to use MiniSAT22~\cite{een2003extensible} in the end.

\subsection{\otfi Guarantees}
\label{sec:otfi:guarantees}

\otfi provides hard security guarantees for a specific fault experiment conducted on the analyzed circuit.
This fault experiment is defined by the security engineer analyzing the circuit in the fault specification and is in line with the threat model of the design.
The fault specification consists of the \textit{(i)} definition of the fault model and the \textit{(ii)} description of the target circuit.

In the fault model~\textit{(i)}, the \otfi user defines the fault capabilities of the attacker, which are specified in the threat model of the analyzed circuit.
This definition comprises the number of faults the attacker can simultaneously inject into the circuit, the effects, and the locations of the faults.
For the fault locations, the security engineer can either target specific gates or instrument \otfi to exhaustively inject faults into all gates of the circuit.
\otfi injects a fault into these targeted gates by replacing the boolean function of the gate according to the fault mapping specified in the fault model.
Here, \otfi supports transient or permanent fault effects.

The target circuit~\textit{(ii)} is the subpart of the overall circuit containing the security-critical logic and the corresponding fault countermeasure the security engineer aims to analyze with \otfi.
This circuit is defined in the fault specification by providing the names of input and output ports of a module or certain gates.
\otfi then automatically extracts the target circuit between these inputs and outputs.
In addition to the names of these ports or gates, the \otfi user needs to specify a specific circuit state, \ie values for the inputs and outputs.

Depending on the configuration, \otfi can reveal whether a fault has \textit{(a)} an unspecific (\cf Section~\ref{sec:otfi:arbitfault}) or \textit{(b)} a specific (\cf Section~\ref{sec:otfi:specfault}) effect.
More concretely, \otfi can formally verify \textit{(a)} whether or not any fault specified in the fault model can change the input-output relation of the target circuit without triggering the fault countermeasures.
Additionally, SYNFI can formally show \textit{(b)} whether or not it is possible to enter a specific circuit state from a given circuit state without triggering the countermeasures using a fault.

Note that SYNFI is designed to detect faults manipulating the input-output relation of the analyzed circuit.
Hence, classes of fault attacks not impacting this relation, \eg safe error~\cite{yen2000checking} or ineffective attacks~\cite{dobraunig2018sifa}, are not in the scope of \otfi.

When the input circuit state space is small, \eg a counter logic, multiple fault experiments for each possible circuit state can be conducted.
Then, \otfi provides comprehensive security guarantees for the analyzed circuit.
For larger circuit state spaces, the security engineer needs to focus on verifying specific states which are particularly security-sensitive or are a representative of the possible states.

\paragraph{False-positive Results.}
\otfi can produce false-positive results when the target circuit is too loosely specified.
As described in Section~\ref{sec:otfi:diffgraph}, the \otfi user does not need to provide the enire input circuit state in the fault specification.
The non-defined inputs provide more freedom to the SAT solver and the solver can freely set these inputs as long as the non-faulty target graph produces the specified output circuit state.

However, in some circuits, the SAT solver could find a circuit state which cannot occur during normal operation. 
Then, a false-positive result is returned, requiring a manual inspection of unexpected effective faults.

Note that the approach of \otfi is to consider states that occur during normal operation and analyze how a fault changes the behavior. 
Using a faulty starting state means analyzing how a fault can change a faulty starting state. 
This is a fault that is beyond the defined fault model and actually corresponds to a stronger fault model.
When a false-positive like this occurs, the security engineer can simply manually exclude it or constrain \otfi more tightly to avoid the need for manual inspection. 
In some cases, the false positive may also provide a hint to the security engineer about faults that can occur with stronger fault models and this can be an input for an extended analysis with a stronger fault model.

Overall, in our analysis, fault positives have not turned out to be a severe limitation as effective faults have occurred rarely in fault-hardened circuits and it was possible to handle them by constraining \otfi more tightly to specific circuit states.

\paragraph{False-negative Results.}
\otfi cannot produce false-negative results within the bounds of the fault specification.
The security engineer only needs to ensure that the fault specification matches the threat model of the analyzed circuit.
For example, when the threat model considers an attacker capable of injecting faults with permanent or transient effects, the fault model also needs to model these faults in the fault mappings.

\section{Analysis of \ot}
\label{sec:otfi:cs}
The \ot chip will be deployed in hostile environments allowing an adversary to gain physical access and attempt to inject faults into the device to break its security.
Therefore, in this chapter, we utilize \otfi to actively contribute to the security of the \ot chip before the tape-out by performing a pre-silicon fault analysis.
As analyzing the entire chip consisting of a wide variety of IP blocks is far beyond the scope of this paper and not all modules actually need to provide fault-resiliency, we selected, together with the \ot project team, the most security-critical modules for our analysis.
In particular, we focused on analyzing \textit{(i)} the unprotected AES module and \textit{(ii)} the protected life cycle controller, the lockstep mode of the CPU, and generic, fault-hardened building blocks.
We utilized \otfi to \textbf{(FE)} reveal the \underline{f}aults' \underline{e}ffect to an unprotected module, to \textbf{(FD)} check whether \underline{f}aults can be \underline{d}etected by the countermeasures, and to \textbf{(FS)} verify that \underline{f}aults cannot enable an adversary to enter a specific \underline{s}tate without triggering the countermeasures.
For all experiments, we injected up to a certain number of simultaneous faults specified in the threat model of each module into the circuit.
Our analysis is conducted on the unmodified netlist synthesized with the internal \ot hardware design flow consisting of the Synopsys DC synthesis tool and a proprietary standard cell library.
\newline\textbf{Results.}
With \otfi, we revealed that the \textit{(i)} AES module is susceptible to single faults enabling an adversary to perform attacks on a round-reduced AES, extract temporary encryption results over the software interface, or hijack the execution-flow of the AES FSM. 
For the other analyzed modules \textit{(ii)}, our analysis showed that they provide adequate protection, \ie all modules can withstand or detect at least single fault attacks.

\subsection{AES}
\label{sec:otfi:cs:aes}
The AES module of \ot is a hardware accelerator providing a secure encryption and decryption mechanism for protocols used by the chip.
As this IP block is one of the most crucial elements of the RoT element, we analyze in detail the behavior of the most security-critical parts of the module when influenced by faults.
In comparison to related work (\cf Section~\ref{sec:otfi:relwork}), we focus on assessing generic hardware primitives, such as state machines and counters, instead of performing specific cryptographic data-flow attacks, such as SIFA~\cite{dobraunig2018sifa} or DFA~\cite{piret2003differential}.
\newline\textbf{Results.}
Our analysis revealed several fault attack vectors for the unprotected AES module.
In particular, \otfi showed that single faults into the AES round counter, handshake signals, and certain FSMs could enable an adversary to break the security of the module.
Based on these verification results, we developed several fault hardening techniques, reassessed their security, and contributed them to the \ot project.

\subsubsection{AES Round Counter}
\label{sec:otfi:cs:aes:ccrc}
The AES~\cite{daemen1999aes} block cipher performs, depending on the mode of operation, a certain number of encryption rounds.
This round counter, which is generated in an FSM, is security-critical, as a fault hijacking the counter value could weaken the cryptographic strength of the AES~\cite{biryukov2004boomerang}.
\begin{table}[]
\centering
\caption{Verification results for the AES round counter performed on a 16-core machine.}
\label{otfi:tab:aesrndctr}
\begin{tabular}{clcccccc}
\hline
 &
  \multicolumn{1}{c}{\textbf{Target}} &
  \textbf{Setting} &
  \textbf{\begin{tabular}[c]{@{}c@{}}Simult.\\ Faults\end{tabular}} &
  \textbf{\begin{tabular}[c]{@{}c@{}}Effective\\ {[}\%{]}\end{tabular}} &
  \textbf{\begin{tabular}[c]{@{}c@{}}Total\\ {[}\#{]}\end{tabular}} &
  \textbf{\begin{tabular}[c]{@{}c@{}}Execution\\ {[}s{]}\end{tabular}} &
  \textbf{\begin{tabular}[c]{@{}c@{}}Circuit\\ {[}GE{]}\end{tabular}} \\ \hline
\ding{172} & \makecell{Unprotected Round Counter} & \textbf{FE} & 1 & 55.56 & 18     & 4.4    & 20.25 \\
\ding{173} & \makecell{Unprotected Round Counter} & \textbf{FS} & 2 & 2.65  & 302    & 4.47   & 20.25 \\
\ding{174} & \makecell{Protected Round Counter}   & \textbf{FD} & 2 & 0.13  & 34,652 & 366.81 & 156   \\ \hline
\end{tabular}
\end{table}
\newline\textbf{Unprotected round counter.}
To analyze the resilience of the round counter against faults, we first \ding{172} utilize \otfi to reveal if the round counter circuit is generally susceptible to faults, \ie it is possible to arbitrarily manipulate the counter value.
Then \ding{173}, we determine how many simultaneous faults are required to manipulate the counter to a specific value.
\begin{lstlisting}[language=json, caption= {Fault specification for the round counter.}, label={lst:otfi:faultmodelrndcntr}]
"Fault Specification":
    "Target Circuit":
        "inputs": ["rnd_ctr_q": "4'b0001"], 
        "outputs": ["rnd_ctr_d": "4'b0010"]
    "Fault Model":
        "Sim. Faults": 1 or 2, "Fault Locations": ["*"]
\end{lstlisting}
To conduct this analysis, we describe the circuit of interest and the fault model in the fault specification file as shown in Listing~\ref{lst:otfi:faultmodelrndcntr}.
We configure \otfi to analyze the logic in between the \texttt{rnd\_ctr} register responsible for incrementing the value and set the input value of the counter circuit to $1$ and the expected output value to $2$.
For the fault model, we instrument \otfi to exhaustively induce one or two simultaneous faults into all available gates of the circuit. 
We provide and describe the used fault model configuration to verify the round counter in more detail in Appendix~\ref{sec:otfi:appendix_fault_model}.

Table~\ref{otfi:tab:aesrndctr} shows the evaluation report generated by \otfi.
The setting column in the table specifies how \otfi considers a fault to be effective.
In the \textbf{(FE)} mode, any \underline{f}ault having an arbitrary \underline{e}ffect to the input-output relation of the circuit is considered to be an effective fault.
In \textbf{(FD)}, an effective \underline{f}ault is a fault manipulating the circuit's output and the fault countermeasures did not \underline{d}etect, \ie did not trigger the alert signal, this fault.
Finally, \textbf{(FS)} refers to a \underline{f}ault changing the output of the circuit to a \underline{s}pecific state without triggering the countermeasures.
Moreover, the table shows the total number of injected faults and the percentage of the effective faults.
The effective fault percentage number indicates how many of the total number of injected faults \otfi considers to be effective.
Finally, the table highlights the execution time of \otfi and the circuit size in gate equivalent~(GE). 
Note that the circuit size refers to the fault affected target circuit extracted by the \otfi framework, which is a subcircuit of the whole circuit.

As shown in the first row \ding{172}, a single fault into the circuit enables a fault attacker to manipulate the round counter value.
To manipulate the round counter to a specific value, \otfi reveals in the second row \ding{173} that an adversary needs to induce at least two simultaneous faults.
\newline\textbf{Hardened round counter.}
To enhance the resilience of the counter against faults, we extend the FSM to generate an up counting (the round counter) and a redundant down counting counter value.
We redundantly instantiate this FSM, combine the generated counters, and add an error logic capable of detecting an ongoing fault attack.
\begin{lstlisting}[style={verilog-style}, caption= {Round counter protection in the \texttt{aes\_cipher\_control} module.}, label={lst:otfi:rnd_ctr_hardening}]
// Instantiate redundant FSMs.
for (genvar i = 0; i < 3; i++) begin : gen_fsm
  aes_cipher_control_fsm u_aes_cipher_control_fsm_i (
    .rnd_ctr_q_i     ( rnd_ctr_q           ),
    .rnd_ctr_d_o     ( mr_rnd_ctr_d[i]     ),
    .rnd_ctr_rem_q_i ( rnd_ctr_rem_q       ),
    .rnd_ctr_rem_d_o ( mr_rnd_ctr_rem_d[i] ),
    ...);
end
// Combine counter signals.
always_comb begin : combine_counter_signals
  for (int i = 0; i < 3; i++) begin
    rnd_ctr_d     |= mr_rnd_ctr_d[i];
    rnd_ctr_rem_d |= mr_rnd_ctr_rem_d[i];
  end
end
// Generate sum.
assign rnd_ctr_sum = rnd_ctr_q + rnd_ctr_rem_q;
assign rnd_ctr_err = (rnd_ctr_sum != num_rounds_q) ? 1'b1 : 1'b0;
\end{lstlisting}
To ensure that the synthesis tool does not weaken the redundancy-based protection mechanism shown in Listing~\ref{lst:otfi:rnd_ctr_hardening}, we reassess its security using \otfi.
In particular, we utilize the framework to evaluate whether the circuit is capable of \underline{d}etecting a single \underline{f}ault arbitrarily manipulating the counter value \textbf{(FD)}.

\otfi could formally verify that, in a specific circuit state, a single fault cannot manipulate the counter value without triggering the alert signal.
This specific circuit state comprises a fixed counter input value of $1$ and a counter output value of $2$, all other non-defined inputs of the circuit are automatically set by the SAT solver \otfi internally uses.
We argue that testing a single circuit state, \ie an input-output pair, is sufficient to verify that the tooling of the design flow does not remove the redundancy-based countermeasures.
For two simultaneous faults, \otfi reveals in Row~\ding{174} in Table~\ref{otfi:tab:aesrndctr}, that at least one fault into the error logic and one fault into the input or output shared round counter register are required to tamper the counter value without raising the alert.

\subsubsection{AES Handshake Signals}
\label{sec:otfi:cs:aes:mubi}
Internally, the AES IP consists of a variety of handshake signals responsible for influencing the data- and control-flow of the encryption.
As manipulating the \texttt{out\_valid\_o} signal would allow an adversary to leak temporary encryption data to the software interface of the AES, we exemplarily focus on analyzing this signal.
More specifically, we instrument \otfi to show whether it is possible to manipulate this signal to a \underline{s}pecific value \textbf{(FS)}, \ie from a logical $0$ to a logical $1$.
Here, we configure \otfi to inject a single fault into the FSM circuit responsible for driving this signal.
The verification result in Row~\ding{172} in Table~\ref{otfi:tab:aessignals} shows that already a single fault induced into the circuit enables an adversary to tamper the handshake signal.
\begin{table}[]
\centering
\caption{Verification results for the AES handshake signal on a 16-core machine.}
\label{otfi:tab:aessignals}
\begin{tabular}{clcccccc}
\hline
 &
  \multicolumn{1}{c}{\textbf{Target}} &
  \textbf{Setting} &
  \textbf{\begin{tabular}[c]{@{}c@{}}Simult.\\ Faults\end{tabular}} &
  \textbf{\begin{tabular}[c]{@{}c@{}}Effective\\ {[}\%{]}\end{tabular}} &
  \textbf{\begin{tabular}[c]{@{}c@{}}Total\\ {[}\#{]}\end{tabular}} &
  \textbf{\begin{tabular}[c]{@{}c@{}}Execution\\ {[}s{]}\end{tabular}} &
  \textbf{\begin{tabular}[c]{@{}c@{}}Circuit\\ {[}GE{]}\end{tabular}} \\ \hline
\ding{172} & Unprotected Handshake Signal & \textbf{FS} & 1 & 83.83 & 31   & 4.49    & 32 \\
\ding{173} & Protected Handshake Signal   & \textbf{FS} & 3 & 15.32 & 8436 & 270.68  & 38 \\ \hline
\end{tabular}
\end{table}
\begin{lstlisting}[style={verilog-style}, caption= {\texttt{out\_valid\_o} signal generation in the \texttt{aes\_cipher\_control\_fsm} module.}, label={lst:otfi:out_valid}]
module aes_cipher_control_fsm (
  output logic             out_valid_o,
  input  logic [3:0]       rnd_ctr_q_i,
  ...
);
  assign num_rounds_regular = num_rounds_q_i - 4'd1;
  unique case (aes_cipher_ctrl_cs)
    ROUND: begin
      advance = (dec_key_gen_q_i | sub_bytes_out_req_i) & key_expand_out_req_i;
      if (advance) begin
        // Are we doing the last regular round?
        if (rnd_ctr_q_i == num_rounds_regular) begin
          if (dec_key_gen_q_i) begin
            out_valid_o = 1'b1;
          ...
\end{lstlisting}
The detailed verification summary reporting the fault-affected cells shows that the adversary can induce faults either \textit{(i)} directly into the \texttt{out\_valid\_o} signal (Line~14 in Listing~\ref{lst:otfi:out_valid}), the \textit{(ii)} comparisons in the output logic (Line~12 in Listing~\ref{lst:otfi:out_valid}), or the \textit{(iii)} control signals (Line~13 in Listing~\ref{lst:otfi:out_valid}) of the FSM.
Hence, to comprehensively protect the handshake signal, we must consider all three attack vectors. 
\newline\textbf{Multi-bit encoding.}
To protect critical handshake signals \textit{(i)}, we extend the AES IP to adopt the multi-bit encoding the \ot project uses in other hardware modules.
\begin{lstlisting}[style={verilog-style}, caption= {Encoded multi-bit signals.}, label={lst:otfi:mubi}] 
typedef enum logic [2:0] { SP2V_HIGH = 3'b011, SP2V_LOW  = 3'b100 } sp2v_e;
\end{lstlisting}
In the multi-bit encoding approach shown in Listing~\ref{lst:otfi:mubi}, a 1-bit signal is encoded into a 3-bit signal resulting in a Hamming distance of three.
Here, the first two bits represent the logical value and the third bit is the inverse of the value to encode.

To verify that the synthesis step does not weaken the security guarantees of multi-bit signals by simplifying the encoding in the optimization phase, we use the \otfi framework to inject faults into the encoded \texttt{out\_valid\_o} signal.
In particular, we use \otfi to reveal if it is possible to manipulate the encoded signal to a \underline{s}pecific value \textbf{(FS)}, \ie from \texttt{SP2V\_LOW} to \texttt{SP2V\_HIGH}.
The verification result in Row~\ding{173} in Table~\ref{otfi:tab:aessignals} confirms the expected security bound of three, \ie \otfi could not find an effective fault when inducing one or two simultaneous faults.
\newline\textbf{Multi-rail FSM.}
As a single fault into the output logic \textit{(ii)} of the FSM, \eg the comparison in Line~12 in Listing~\ref{lst:otfi:out_valid}, is enough to tamper the out valid signal, we design and deploy a redundant multi-rail FSM scheme.

\begin{figure*}[h]
  \centering
  \includegraphics[width=0.6\textwidth]{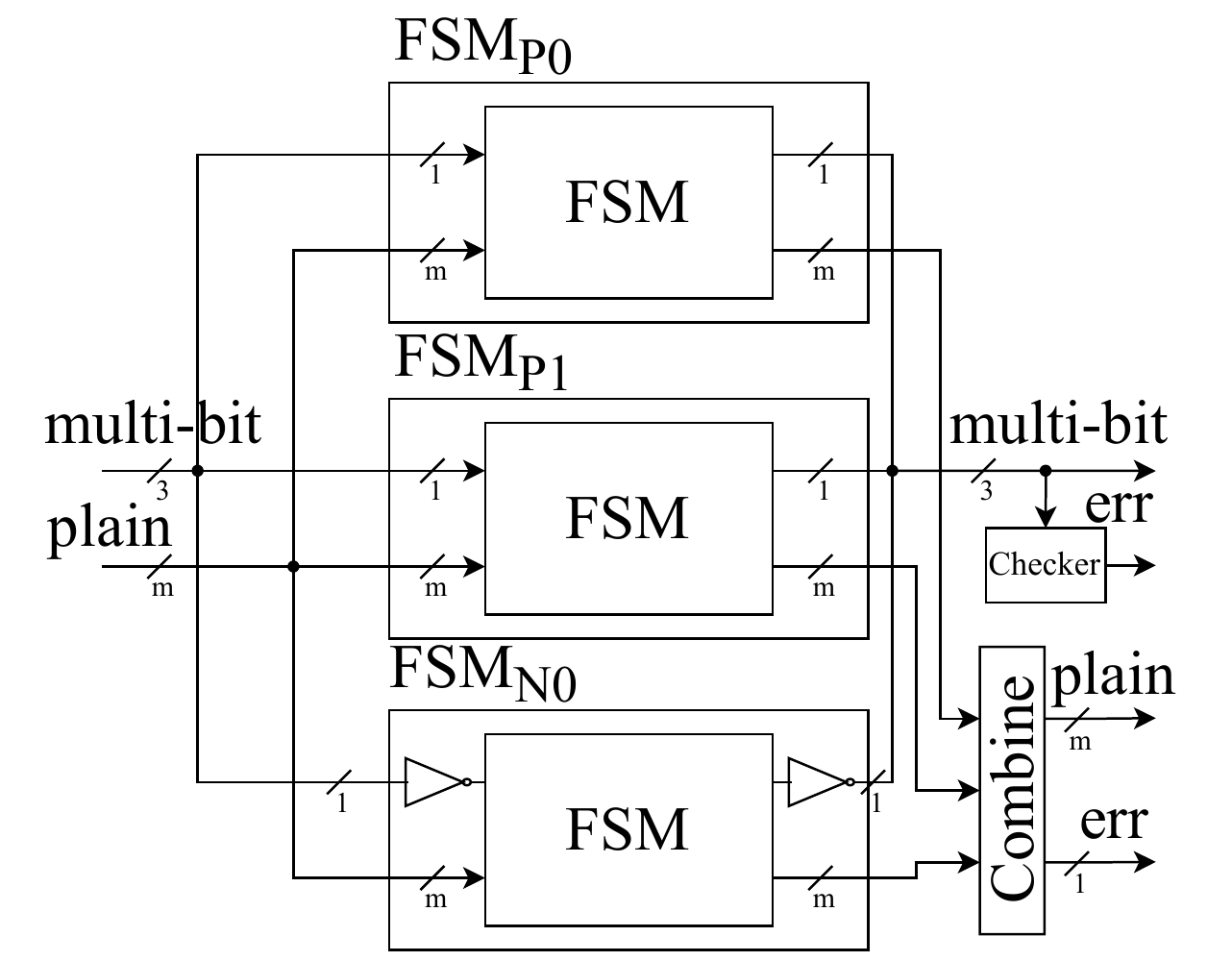}
  \caption{Multi-rail FSM approach.}
  \label{fig:otfi:multirail}
\end{figure*}

The multi-rail scheme, as shown in Figure~\ref{fig:otfi:multirail}, instantiates the unmodified $FSM$ in a triple modular redundancy mode. 
Unencoded m-bit signals are independently processed by the three state machines and an output logic is responsible for combining and checking the resulting signals.
On a comparison mismatch, an alert signal is triggered.
For multi-bit encoded signals (\cf Listing~\ref{lst:otfi:mubi}), the first two bits are processed by the positive $FSM_{P0}$ and $FSM_{P1}$ rail.
The inverted third bit is processed by the negative $FSM_{N0}$ rail.
Combining these signals at the output again produces an encoded multi-bit signal.
If a fault in one or two FSMs modifies the signal, an invalid code word is produced, which is detectable by a checker unit.

\begin{lstlisting}[style={verilog-style}, caption= {Fault resistant multi-rail FSM in the \texttt{aes\_cipher\_control} module.}, label={lst:otfi:multi_rail},float,floatplacement=H]
assign sp_dec_key_gen_q = {dec_key_gen_q}
// For every bit in the Sp2V signals, one separate rail is instantiated.
for (genvar i = 0; i < 3; i++) begin : gen_fsm
  if (SP2V_LOGIC_HIGH[i] == 1'b1) begin : gen_fsm_p
    aes_cipher_control_fsm_p u_aes_cipher_control_fsm_i (
      .rnd_ctr_q_i     ( rnd_ctr_q           ),
      .num_rounds_q_i  ( num_rounds_q        ),
      .dec_key_gen_q_i ( sp_dec_key_gen_q[i] ),
      .out_valid_o     ( sp_out_valid[i]     ),
      ...
    );
  end else begin: gen_fsm_n
    aes_cipher_control_fsm_n u_aes_cipher_control_fsm_i (
      .rnd_ctr_q_i     ( rnd_ctr_q           ),
      .num_rounds_q_i  ( num_rounds_q        ),
      .dec_key_gen_q_i ( sp_dec_key_gen_q[i] ),
      .out_valid_o     ( sp_out_valid[i]     ),
      ...
    );
  end
end
// Convert sparsified outputs to sp2v_e type.
assign out_valid_o = sp2v_e'(sp_out_valid);
\end{lstlisting}
Listing~\ref{lst:otfi:multi_rail} shows the multi-rail FSM approach integrated into the AES module.
In this approach, three redundant FSMs are instantiated where the two \texttt{aes\_cipher\_control\_fsm\_p} FSMs produce a positive output and the \texttt{aes\_cipher\_control\_fsm\_n} FSM a negated output.
Combined, they form a multi-bit signal with a Hamming distance of three.
As the multi-rail approach requires to instantiate FSMs redundantly, the area increases from $211.15\,GE$ for the \texttt{aes\_cipher\_control\_fsm\_p} FSM to $908.81\,GE$ for the whole \texttt{aes\_cipher\_control} module including the redundant FSMs, the combination logic, as well as other countermeasures, such as the counter error logic introduced in Section~\ref{sec:otfi:cs:aes:ccrc}.
\begin{table}[H]
\centering
\caption{Verification results for the AES multi-rail FSM on a 16- or 72-core* machine.}
\label{otfi:tab:aesmultirail}
\begin{tabular}{clcccccc}
\hline
 &
  \multicolumn{1}{c}{\textbf{Target}} &
  \textbf{Setting} &
  \textbf{\begin{tabular}[c]{@{}c@{}}Simult.\\ Faults\end{tabular}} &
  \textbf{\begin{tabular}[c]{@{}c@{}}Effective\\ {[}\%{]}\end{tabular}} &
  \textbf{\begin{tabular}[c]{@{}c@{}}Total\\ {[}\#{]}\end{tabular}} &
  \textbf{Execution} &
  \textbf{\begin{tabular}[c]{@{}c@{}}Circuit\\ {[}GE{]}\end{tabular}} \\ \hline
\ding{172} & \makecell{Multi-Rail FSM\\ loose config}  & \textbf{FS} & 1 & 2.46 & 122        & 8.2\,s    & 96.5   \\
\ding{173} & \makecell{Multi-Rail FSM\\ tight config}  & \textbf{FS} & 1 & 0    & 573        & 266.87\,s & 355.75 \\ 
\ding{174} & \makecell{Multi-Rail FSM\\ tight config}  & \textbf{FS} & 2 & 0    & 170,982    & 1.01\,h   & 355.75 \\
\ding{175} & \makecell{*Multi-Rail FSM\\ tight config} & \textbf{FS} & 3 & 0.02 & 35,222,293 & 38.27\,h  & 355.75 \\ \hline
\end{tabular}
\end{table}

To verify that the synthesis tool does not remove the redundant FSMs, we utilize \otfi to analyze the resilience of the multi-rail approach against faults.
In particular, we instrument the framework to reveal whether there exists a \underline{f}ault enabling an adversary to manipulate the \texttt{out\_valid\_o} signal to a \underline{s}pecific value \textbf{(FS)}, \ie from \texttt{SP2V\_LOW} to \texttt{SP2V\_HIGH}, without triggering the fault countermeasure.
\begin{lstlisting}[language=json, caption= {Fault specification for the multi-rail FSM.}, label={lst:otfi:faultmodelmultirailfsm},float,floatplacement=H]
"Fault Specification":
    "Target Circuit":
        "inputs": ["rnd_ctr": "2"], 
        "outputs": ["out_valid_o": "SP2V_LOW"],
        "expected fault outputs": ["out_valid_o": "SP2V_HIGH"],
        "alerts": ["rnd_ctr_err": "0"]
\end{lstlisting}
Using the fault specification file shown in Listing~\ref{lst:otfi:faultmodelmultirailfsm}, \otfi automatically extracts the circuit between the round counter register and the out valid signal, \ie the multi-rail FSM including the input, output, and error logic.
By setting the counter value to $2$, we force the circuit in a state where the out valid signal is set to a logical $0$ in the fault-free setting.
As shown in Row~\ding{172} in Table~\ref{otfi:tab:aesmultirail}, in this \otfi configuration, already a single fault can be sufficient to produce an effective fault, \ie the out valid signal is set to a logical $1$ and the error was not triggered.
The framework reports that all of these four effective faults occur when faulting the \texttt{num\_rounds} register value.
Since the value of this register is used for the comparison in Line~12 in Listing~\ref{lst:otfi:out_valid} in all redundant FSMs, the out valid signal is set to a logical 1.
Nevertheless, this verification result can only provide a limited statement about the security of the multi-rail approach as the \otfi framework was minimally constrained.
By only defining the input value of the round counter register, the SAT solver is \textit{loosely} constrained (\cf Section~\ref{sec:otfi:guarantees}) and automatically sets the \texttt{dec\_key\_gen\_q\_i} and \texttt{advance} signals in the boolean formula of the differential graph to a logical $1$.
Setting these variables is possible, as in the non-faulty reference circuit the \texttt{out\_valid\_o} always stays at \texttt{SP2V\_LOW} when the round counter value is $2$.
Hence, as all redundant FSMs set the \texttt{out\_valid\_o} to the same value, the error signal is not set.

To avoid these false-positive results, we more \textit{tightly} configure \otfi by further defining the inputs $dec\_key\_gen\_q\_i = 0$ and $key\_expand\_out\_req\_i = 0$ in the fault specification file.
In this configuration, the extracted circuit \otfi analyzes increases from $96.5\,GE$ to $355.75\,GE$, as the tool finds more paths from the defined inputs to the outputs. 
Now, as expected and depicted in Row~\ding{173} and \ding{174} in Table~\ref{otfi:tab:aesmultirail}, one or two simultaneous faults cannot manipulate the out valid signal.
Starting with three simultaneous faults into the circuit, we observe effective faults (\cf Row~\ding{175} in Table~\ref{otfi:tab:aesmultirail}).
These effective faults manipulating the out valid signal are caused by inducing bit-flips into variables used by the redundant FSMs.
To demonstrate the possibility of scaling \otfi to the cloud and as the number of possible fault combinations for three simultaneous faults, \ie fault location and fault mapping, explodes, we conducted this experiment on a 72-core server. 
We measured a total run time of $38.27\,h$ and a maximum memory consumption of less than $8\,GB$ for injecting $35,222,292$ faults into the circuit of a size of $355.75\,GE$.
\newline\textbf{Shadow registers.}
As discussed in the initial experiment in Section~\ref{sec:otfi:cs:aes:mubi}, handshake signals also can be tampered by faulting \textit{(iii)} control signals used by the FSM.
To protect security-critical control signals, which are provided by the software over a register interface, the AES modules stores them in dedicated shadow registers.
These registers constantly compare the two values and on a comparison mismatch caused by, for example a fault, an alert is raised forcing the AES module in a terminal state.
\subsubsection{Sparsely Encoded State Machines}
\label{sec:otfi:cs:aes:fsmencoding}
Finite-state machines are, as seen in Section~\ref{sec:otfi:cs:aes:mubi}, security-critical hardware elements as they are responsible for setting control signals used by the data path.
\begin{lstlisting}[style={verilog-style}, caption= {Finite-state machine with a state encoding vulnerable to faults.}, label={lst:otfi:fsm}]
typedef enum logic [3:0] { IDLE, INIT, ROUND, ..., ERROR } aes_cipher_ctrl_e;
always_comb begin : aes_cipher_ctrl_fsm
  unique case (aes_cipher_ctrl_cs)
    IDLE: begin
      if <condition>:
        aes_cipher_ctrl_ns = INIT;
      end
    end
    INIT: begin
    ...
\end{lstlisting}
In a state machine, the next-state logic derives the next state from the current state and a set of inputs.
As seen in Listing~\ref{lst:otfi:fsm}, the state variable stored in the state register is typically represented as a simple enum.
However, as the minimum Hamming distance between two states is $1$, a single fault into the state registers would allow an adversary to hijack the control-flow of the FSM, \ie skip or enter a normally non-reachable state.
  
In order to mitigate this threat, we deploy the sparse FSM state encoding technique used by different \ot modules into the AES.
\begin{lstlisting}[style={verilog-style}, caption= {Sparsely encoded FSM state.}, label={lst:otfi:state_encoding},float,floatplacement=H]
typedef enum logic [5:0] {
  IDLE        = 6'b001001, INIT        = 6'b100011, ROUND       = 6'b111101,
  FINISH      = 6'b010000, PRNG_RESEED = 6'b100100, CLEAR_S     = 6'b111010,
  CLEAR_KD    = 6'b001110, ERROR       = 6'b010111
} aes_cipher_ctrl_e;
\end{lstlisting}
The encoding, which is shown in Listing~\ref{lst:otfi:state_encoding}, assures a minimum Hamming distance between the states of $3$, increasing the resistance against faults.
Additionally, we introduce a default error state, which is entered when the state value does not match the \texttt{aes\_cipher\_ctrl\_e} enum.
Now, if a fault flips bits in the state variable, with a high probability, the terminal error state is entered.
\begin{table}[]
\centering
\caption{Verification results for the AES FSM state encoding on a 16-core machine.}
\label{otfi:tab:aesstateenc}
\begin{tabular}{lcccccc}
\hline
  \multicolumn{1}{c}{\textbf{Target}} &
  \textbf{Setting} &
  \textbf{\begin{tabular}[c]{@{}c@{}}Simult.\\ Faults\end{tabular}} &
  \textbf{\begin{tabular}[c]{@{}c@{}}Effective\\ {[}\%{]}\end{tabular}} &
  \textbf{\begin{tabular}[c]{@{}c@{}}Total\\ {[}\#{]}\end{tabular}} &
  \textbf{\begin{tabular}[c]{@{}c@{}}Execution\\ {[}s{]}\end{tabular}} &
  \textbf{\begin{tabular}[c]{@{}c@{}}Circuit\\ {[}GE{]}\end{tabular}} \\ \hline
  Encoded FSM states & \textbf{FS} & 3 & 15 & 29  & 16.09 & 90.5  \\ \hline
\end{tabular}
\end{table}

To verify that an aggressive synthesis setting does not reduce the security by altering the state encoding, we utilize \otfi to analyze the \texttt{aes\_cipher\_control\_fsm} FSM.
In particular, we determine, how many faults are required to hijack the control-flow of the FSM by skipping a certain state and directly enter a normally not reachable \underline{s}tate, \ie \textbf{(FS}).
For this experiment, we instrument \otfi to analyze the next-state logic and to inject faults directly into the state registers.
\otfi shows that one or two simultaneous bit-flips into the state registers triggers the alert signal, \ie the FSM enters the error state.
When inducing three simultaneous faults, as shown in Table~\ref{otfi:tab:aesstateenc}, the attacker is able to redirect the control-flow of the FSM.
\newline\textbf{FSM optimizations.}
Several synthesis tools also apply optimization passes to state machines.
Yosys, for example, removes unused control signals, merges states, and recodes the FSM state variables stored in the state registers~\cite{YosysManual}.
To analyze the impact of these optimization on the security of the sparsely encoded states, we synthesize the \texttt{aes\_cipher\_control\_fsm} module with Yosys using an aggressive optimization strategy.
Similar to the previous experiment, we configure \otfi to skip a state and directly enter a typically non-reachable state.
Our result shows that, in comparison to the verification in Table~\ref{otfi:tab:aesstateenc}, now 2 instead of 3 simultaneous faults are already sufficient to skip the FSM state, \ie the FSM optimization weakens the encoding.
To prevent these optimizations, Yosys can be parameterized with the \texttt{nofsm} flag.
In summary, this experiment shows that synthesis optimizations configured by different stakeholders, \eg trying to minimize the area of the design, could have fatal security implications.

\subsection{Life Cycle Controller}
\label{sec:otfi:cs:lc}
\ot can be transferred into different operational states depending on where the device is deployed, \ie at the customer or the manufacturer.
The state switching mechanism is implemented in hardware in the life cycle controller module.
As certain states, \eg the return material authorization~(RMA) state, enable security-sensitive features, such as access to the debug port, the life cycle controller is hardened against fault attacks.
\newline\textbf{Results.}
\otfi verified that the analyzed fault-hardened primitives of the life cycle controller provide adequate protection.
Specifically, we could confirm that the countermeasures prevent the exploitation of single or double faults for all critical attack vectors.

\subsubsection{Entering the RMA State}
\label{sec:otfi:cs:rma}
The core mechanism of the life cycle controller IP is an FSM responsible for determining the life cycle state of \ot.
To hinder an adversary from entering the security-sensitive RMA state, which is used to debug the RoT chip when returned to the manufacturer, this state transition is only permitted when possessing a 128-bit unlock token.
Internally, the state machine checks the validity of the token in three different FSM states.
This redundancy-based mechanism and the state encoding technique guaranteeing a minimum Hamming distance of $7$ between the state symbols form the fault protection strategy of the controller.
Based on this description of the fault-hardening mechanisms, we identified two major attack vectors for a fault attacker: \textit{(i)} glitch the token comparisons three times or \textit{(ii)} hijack the execution of the FSM by glitching the state symbols.
\begin{table}[]
\centering
\caption{Verification results for entering the RMA state on a 16-core machine.}
\label{otfi:tab:lcrma}
\begin{tabular}{clcccccc}
\hline
 &
  \multicolumn{1}{c}{\textbf{Target}} &
  \textbf{Setting} &
  \textbf{\begin{tabular}[c]{@{}c@{}}Simult.\\ Faults\end{tabular}} &
  \textbf{\begin{tabular}[c]{@{}c@{}}Effective\\ {[}\%{]}\end{tabular}} &
  \textbf{\begin{tabular}[c]{@{}c@{}}Total\\ {[}\#{]}\end{tabular}} &
  \textbf{\begin{tabular}[c]{@{}c@{}}Execution\\ {[}s{]}\end{tabular}} &
  \textbf{\begin{tabular}[c]{@{}c@{}}Circuit\\ {[}GE{]}\end{tabular}} \\ \hline
\ding{172} & \makecell{Token comparison\\ in \texttt{TokenHashSt}}   & \textbf{FS} & 1 & 0   & 313 & 32.56 & 215    \\
\ding{173} & \makecell{Token comparison\\ in \texttt{TokenCheck0St}} & \textbf{FS} & 1 & 0   & 349 & 35.87 & 248.25 \\
\ding{174} & \makecell{Token comparison\\ in \texttt{TokenCheck1St}} & \textbf{FS} & 1 & 0   & 310 & 28.26 & 204.75 \\
\ding{175} & Skip token check states                              & \textbf{FS} & 7 & 100 & 7   & 17.44 & 214.5  \\ \hline
\end{tabular}
\end{table}
\newline\textbf{Glitching the comparisons.}
Glitching the three token checks requires a strong adversary capable of injecting three faults in three clock cycles.
We utilize \otfi to test whether these three comparisons are susceptible to a single fault each.
For this verification, we configure \otfi to analyze if it is possible to induce a \underline{f}ault into the next-state logic of the FSM changing the next valid \underline{s}tate \textbf{(FS)}.
In the fault specification file, we instrument the tool to exhaustively induce a single fault into all gates of the next-state logic for each of the three fault experiments.
Row~\ding{172}-\ding{174} in Table~\ref{otfi:tab:lcrma} shows that \otfi could not find a single fault allowing the attacker to enter the next state (from \texttt{TokenHashSt} to \texttt{TokenCheck0St}, $\cdots$) without possessing the required token.
\newline\textbf{Skip the token check states.}
In order to verify that the synthesis step does not weaken the 16-bit FSM state encoding, we utilize \otfi to check whether it is possible to induce \underline{f}aults into the state register allowing the attacker to directly enter the RMA \underline{s}tate \textbf{(FS)}.
Here, we configure \otfi to analyze the FSM and to inject 1 to 7 simultaneous faults into the state register.
As shown in Row~\ding{175} in Table~\ref{otfi:tab:lcrma}, at least $7$ simultaneous faults are required to enter the target state.
This matches the security expectation, \ie a Hamming distance of $7$ for the state symbol encoding.

\subsubsection{Flash Erase Mechanism}
\label{sec:otfi:cs:flash}
Before entering RMA, the life cycle controller erases the flash to hinder an adversary from accessing previously created data. 
Although entering RMA only is possible when knowing a secret, device dependent token, this hardware-backed flash erasing mechanism is meant to be a second line of defense.
Internally, the flash erasing command is directly triggered in the FSM of the life cycle controller.
To ensure that the flash was erased before entering RMA, the acknowledgment sent by the flash controller is checked three times in the FSM.
If one of the acknowledgements was not received, \eg due to a fault, the FSM remains in the current state.
\newline\textbf{Glitching the Encoded Flash Handshake Signal.}
An attacker with access to a valid RMA token aiming to bypass the flash erasing mechanisms needs to suppress the flash erasing command as well as the acknowledgement signal or the corresponding check three times.
However, as the initiate and acknowledgement signal is encoded with a Hamming distance of $4$, the adversary theoretically needs to flip 4-bits four times.
To confirm this behavior, \ie the synthesis tool did not tamper the encoding, we exemplary analyze the resilience of the flash erase initiate signal.
In particular, we instrument \otfi to reveal whether it is possible to induce \underline{f}aults manipulating this signal to a \underline{s}pecific value \textbf{(FS)}, \ie from an encoded $On=4'b1001$ to an encoded $Off=4'b0110$.
This analysis showed that already two simultaneous faults injected into the encoded signal allow an adversary to hijack the initiate signal.
Since bit 0 and 3 as well as 1 and 2 in the encoding are always the same, the synthesis tool decided to merge these signals and only instantiate two instead of four registers driving the bits of the signal.
Hence, the security of the encoding is reduced to a Hamming distance of $2$.

In order to prevent that the four registers instantiated in the HDL code are merged in the synthesized netlist, we augmented the design flow with the \texttt{set\_dont\_touch} parameter.
Now, as shown in Row~\ding{175} in Table~\ref{otfi:tab:lcrma}, the encoding works as expected and an attacker needs at least four simultaneous faults to tamper the encoded signal.

\subsubsection{Locking Mechanism}
\ot limits the number of state transitions and transition attempts to $24$. 
Once this number is reached, the life cycle controller rejects further attempts, effectively locking the device into its current state.
In the life cycle hardware IP block, the counter increment is conducted in the \texttt{lc\_ctrl\_state\_transition} module and the counter value is programmed into the one time programmable~(OTP) memory of \ot in the \texttt{lc\_ctrl\_fsm} FSM.
Hence, an adversary aiming to increase the number of state transitions attempts either needs to fault the counter increment \textit{(i)} or the programming \textit{(ii)} of the value into the OTP. 
\begin{table}[]
\centering
\caption{Verification results for glitching the locking mechanism performed on a 16- or 72-core* machine.}
\label{otfi:tab:lclock}
\begin{tabular}{clcccccc}
\hline
 &
  \multicolumn{1}{c}{\textbf{Target}} &
  \textbf{Setting} &
  \textbf{\begin{tabular}[c]{@{}c@{}}Simult.\\ Faults\end{tabular}} &
  \textbf{\begin{tabular}[c]{@{}c@{}}Effective\\ {[}\%{]}\end{tabular}} &
  \textbf{\begin{tabular}[c]{@{}c@{}}Total\\ {[}\#{]}\end{tabular}} &
  \textbf{Execution} &
  \textbf{\begin{tabular}[c]{@{}c@{}}Circuit\\ {[}GE{]}\end{tabular}} \\ \hline
\ding{172} & Prevent counter incr.   & \textbf{FS} & 1 & 17.81 & 853       & 261.76\,s & 1170   \\
\ding{173} & Reset counter value*    & \textbf{FS} & 3 & 0     & 1,000,000 & 7\,h      & 1170   \\
\ding{174} & Skip \texttt{CntProgSt} & \textbf{FS} & 7 & 100   & 7         & 17.53\,s  & 228.75 \\ \hline
\end{tabular}
\end{table}
\newline\textbf{Skip the Counter Increment.}
Inside the \texttt{lc\_ctrl\_state\_transition} module, an FSM is responsible for updating the counter value.
\begin{lstlisting}[style={verilog-style}, caption= {Life cycle controller counter increment FSM.}, label={lst:otfi:lc_cntr},float,floatplacement=H]
module lc_ctrl_state_transition (
  input  lc_cnt_e           lc_cnt_i,
  output lc_cnt_e           next_lc_cnt_o,
  ...
);
  unique case (lc_cnt_i)
    LcCnt0:   next_lc_cnt_o = LcCnt1;
    LcCnt1:   next_lc_cnt_o = LcCnt2;
    ...
    LcCnt23:  next_lc_cnt_o = LcCnt24;
endcase 
\end{lstlisting}
As illustrated in Listing~\ref{lst:otfi:lc_cntr}, this FSM uses a strong state encoding technique to mitigate fault attacks.
Each state of type \texttt{lc\_cnt\_e} consists of 384-bit with a Hamming distance of $269$ between \texttt{LcCnt0} and \texttt{LcCnt24} and a Hamming of $3$ between \texttt{LcCnt23} and \texttt{LcCnt24}.
To verify that the synthesis flow does not weaken the encoding, we utilize \otfi to verify that a \underline{f}ault cannot manipulate the \texttt{next\_lc\_cnt\_o} variable to a \underline{s}pecific value \textbf{(FS)}.
More concretely, we aim to bypass the counter increment from $LcCnt23$ to $LcCnt24$.
In this scenario, the attacker aims to avoid that the counter increments to the final $LcCnt24$ value and locks the life cycle controller.

As shown in Row~\ding{172} in Table~\ref{otfi:tab:lclock}, already a single fault allows the adversary to avoid that the counter is incremented.
Supported by the generated analysis results, we were able to track back the single point of failure responsible for enabling an attacker flipping the three bits, \ie the Hamming distance between \texttt{LcCnt23} and \texttt{LcCnt24}, with a single fault.
Since the three gates driving the three targeted bits of the \texttt{next\_lc\_cnt\_o} output port are driven by a single gate, attacking this gate or drivers of this gate allow an adversary to manipulate the output counter value.
However, as the counter increment is only prevented once, an attacker only could initiate one additional state transition, making this attack impractical in reality.

Since resetting the counter value to \texttt{LcCnt0} enables the attacker to initiate more additional state transitions, the encoding is also stronger, \ie a Hamming distance of $269$ between \texttt{LcCnt23} and \texttt{LcCnt0}.
To ensure that this strong encoding between these two counter values is correct after the synthesis, we test with \otfi whether it is possible to switch to the \underline{s}pecific \texttt{LcCnt0} value from \texttt{LcCnt23} with \underline{f}aults \textbf{(FS)}.
Similar to the previous experiment, we configure the framework to exhaustively inject 1, 2, and 3 simultaneous faults into the circuit responsible for incrementing the counter value.
Since the possible fault combinations explode, we limited the number of injected faults to $1\,M$ and performed the experiment on a 72-core server in the cloud.
Within this fault threat model, \otfi could not find a single, effective fault, as shown in Row~\ding{173} in Table~\ref{otfi:tab:lclock}.
\newline\textbf{Prevent the Programming of the Counter.}
The programming of the counter value into OTP is initiated in the \texttt{CntProgSt} state in the life cycle controller FSM.
We utilize \otfi to validate that an adversary cannot bypass this state and directly switch to the next state \textbf{(FS)}.
As shown in Row~\ding{174} in Table~\ref{otfi:tab:lclock}, at least $7$ faults, \ie the Hamming distance the encoding guarantees, are required to skip the state.

\subsubsection{Life Cycle Escalation Signal}
\label{sec:otfi:cs:escalation}
When the life cycle controller detects an ongoing attack, the 4-bit encoded \texttt{lc\_escalate\_en} signal is triggered.
This signal is consumed by other hardware modules, \eg the AES IP, and transfers them into an non-escapable error state.
To validate that the optimization passes in the synthesis does not weaken the encoding of the signal, we inject faults into the escalation signal driven in the \texttt{lc\_ctrl\_signal\_decode} module.
In the fault model used by \otfi, we set the input value to $lc\_escalate\_en = On = 4'b1001$, the expected output value to $lc\_escalate\_en\_o = On = 4'b1001$, and the fault output to $lc\_escalate\_en\_o = Off = 4'b0110$ \textbf{(FS)}.
Without constraining the synthesis flow with the \texttt{set\_dont\_touch} parameter (\cf Section~\ref{sec:otfi:cs:flash}), the security of the encoding is reduced to a Hamming distance of $2$, as Synopsys removes the redundant flip-flops.
When applying this constraint to the \texttt{lc\_escalate\_en} flip-flop, at least four simultaneous faults are required to suppress the escalation signal.

\subsubsection{Privilege Escalation in the PROD State}
When \ot is shipped to the customer, the device is put into the PROD state.
In this state, certain features are activated, such as the CPU, and security-critical features, such as debug functionalities, are disabled.
Instead of directly hijacking the life cycle state of \ot, an adversary also could aim to switch on such features in the PROD state.
All features are activated in the \texttt{lc\_ctrl\_signal\_decode} module by setting the corresponding signal from \texttt{Off} to \texttt{On}.
This signal then is transmitted to the corresponding hardware module responsible for activating or deactivating the feature.
Similar to the escalation signal described in Section~\ref{sec:otfi:cs:escalation}, the \ot project uses a 4-bit encoding technique with a Hamming distance of $4$ between \texttt{Off} and \texttt{On}.
For the fault verification of the encoded signal, we configure the input to $lc\_hw\_debug\_en = Off = 4'b0110$, the expected output value to $lc\_hw\_debug\_en\_o = Off = 4'b0110$, and the fault output to $lc\_hw\_debug\_en\_o = On = 4'b1001$ \textbf{(FS)}.
Similar to the previous experiments, the \texttt{set\_dont\_touch} constraint needs to be applied to the registers responsible for driving the \texttt{lc\_hw\_debug\_en} signal to maintain a Hamming distance of $4$.
Then, at least four simultaneous faults are required to enable the debug mode.
While a transient fault only can active the debug functionality for a brief moment, a permanent stuck-at fault could allow an adversary to enable this feature permanently.

\subsection{Ibex}
\label{sec:otfi:cs:ibex}
The RISC-V Ibex processor is the core element of the \ot chip.
In this section, we utilize \otfi to analyze the behavior of the CPU when injecting faults.
To demonstrate the ability of \otfi to handle different netlists, we, contrary to the previous verification setups, analyze the netlist synthesized with the open-source Yosys synthesis tool and the open Nangate45 cell library.
\newline\textbf{Results.}
Our analysis showed that the error logic of the Ibex lockstep mode is capable of detecting a fault into the program counter.

\begin{table}[]
\centering
\caption{Verification results for the Ibex processor on a 16-core machine.}
\label{otfi:tab:ibxresults}
\begin{tabular}{lccccccc}
\hline
 &
  \textbf{Target} &
  \textbf{Setting} &
  \textbf{\begin{tabular}[c]{@{}c@{}}Simult.\\ Faults\end{tabular}} &
  \textbf{\begin{tabular}[c]{@{}c@{}}Effective\\ {[}\%{]}\end{tabular}} &
  \textbf{Total} &
  \textbf{\begin{tabular}[c]{@{}c@{}}Execution\end{tabular}} &
  \textbf{\begin{tabular}[c]{@{}c@{}}Circuit\\ {[}GE{]}\end{tabular}} \\ \hline
 \ding{172} & Glitch the PC  & \textbf{FE} & 1 & 78.1 & 557     & 185.07\,s & 488    \\
 \ding{173} & Glitch the PC  & \textbf{FS} & 2 & 0.02 & 309,500 & 0.72\,h   & 488    \\
 \ding{174} & Lockstep mode  & \textbf{FS} & 1 & 6.31 & 111     & 10.22\,s  & 165.34 \\\hline
\end{tabular}
\end{table}

\subsubsection{Glitching the Program Counter}
\label{sec:otfi:cs:pcglitch}
A fault into the program counter~(PC) allows an attacker to arbitrarily redirect the control-flow of the program executed on the processor~\cite{timmers2016controlling}.
We utilize \otfi to \ding{172} analyze whether the instruction fetch stage of the Ibex is generally susceptible to a fault arbitrarily changing the PC \textbf{(FE)}.
Row~\ding{172} in Table~\ref{otfi:tab:ibxresults} shows that a single fault already is sufficient to manipulate the PC and to redirect the control-flow.
Although targeting a \texttt{NOP} slide does not require an adversary to accurately manipulate the PC, randomly glitching the PC makes it hard for the attacker to exploit the induced fault.
Therefore, we analyze \ding{173} if it is possible to change the program counter to a \underline{s}pecific PC, \ie from the boot address to 32'h40400, using a \underline{f}ault \textbf{(FS)}.
The analysis of \otfi in Row~\ding{173} in Table~\ref{otfi:tab:ibxresults} shows that, with two simultaneously induced faults, glitching the PC to a specific value is hard.
More specifically, for this fault specification, \otfi shows that only $62$ ($0.02\,\%$) out of $309,500$ injected faults manipulate the program counter to the specified value.

\subsubsection{Lockstep Mode}
To protect the execution of software on Ibex from faults, \ot instantiates the CPU redundantly in a dual-core lockstep mode.
In this approach, the input used for the main core is delayed, provided to the redundant core, and the delayed output is compared to the output of the main core.
On a mismatch, a hardware monitor raises an alert.
Similar to the verification setup in Section~\ref{sec:otfi:cs:pcglitch}, we consider an adversary aiming to redirect the control-flow by glitching the program counter.
For the verification, we assume that the attacker already managed to flip a bit in the instruction address generated by the main core but not in the shadow core.
As the error detection logic should raise an error due to the mismatch, a fault attacker needs to additionally inject a fault into this error detection circuit.
\otfi reveals that \textit{(i)} the error detection logic actually raises an error, \ie the synthesis tool did not remove the redundant logic, and that \textit{(ii)} one fault could enable an attacker to suppress the error flag \textbf{(FS)}, as shown in Row~\ding{174} in Table~\ref{otfi:tab:ibxresults}.

\subsection{Generic Primitives}
\label{sec:otfi:cs:prim}
The \ot project provides a set of generic hardware building blocks which are reused in several hardware modules. 
In this section, we analyze the fault protected generic primitives, \ie the counter and the LFSR, using \otfi.
\newline\textbf{Results.}
Our analysis with \otfi confirms that the inspected primitives provide the expected security, \ie a single fault into the protected counter or the LFSR triggers the alert signal of the countermeasures.

\begin{table}[b]
\centering
\caption{Verification results for the \texttt{prim\_double\_lfsr} and \texttt{prim\_count} modules.}
\label{otfi:tab:genericresults}
\begin{tabular}{lccccccc}
\hline
 &
  \textbf{Target} &
  \textbf{Setting} &
  \textbf{\begin{tabular}[c]{@{}c@{}}Simult.\\ Faults\end{tabular}} &
  \textbf{\begin{tabular}[c]{@{}c@{}}Effective\\ {[}\%{]}\end{tabular}} &
  \textbf{Total} &
  \textbf{\begin{tabular}[c]{@{}c@{}}Execution\\ {[}s{]}\end{tabular}} &
  \textbf{\begin{tabular}[c]{@{}c@{}}Circuit\\ {[}GE{]}\end{tabular}} \\ \hline
  \ding{172} & \texttt{prim\_count}           & \textbf{FD} & 2 & 10.82 & 1552    & 37.46   & 29.75 \\
  \ding{173} & \texttt{prim\_double\_lfsr}    & \textbf{FD} & 2 & 0.05  & 7827    & 74.44   & 116.75 \\
  \ding{174} & \texttt{prim\_double\_lfsr}    & \textbf{FD} & 3 & 0.09  & 340,692 & 2114.69 & 116.75   \\\hline
\end{tabular}
\end{table}

\subsubsection{Counter}
\label{sec:otfi:cs:prim_counter}

The \texttt{prim\_count} module provides a fault protected generic counter primitive which is used by different modules.
This module offers a flexible parameterization interface allowing the hardware designer to define the mode, the start value, and the bit width of the counter.
In order to mitigate faults manipulating the counter value, the \texttt{prim\_count} module implements the double count or cross count protection mode.
While in the double count mode two redundant counters are compared, in the cross count mode the values of the up counting counter and the down counting counter are added and compared to a constant.
On a comparison mismatch, a fault is detected and an error is triggered.

To ensure that the synthesis does not remove the redundant counter, we use the \otfi framework to test the resilience of the module against faults.
In particular, we check whether the countermeasure can \underline{d}etect a \underline{f}ault arbitrarily changing the output of the counter value \textbf{(FD)}.
For this, we configure \otfi to inject faults into the counter logic, the counter registers, and the comparison logic.
With this description, \otfi automatically extracts the target circuit~($29.75\,GE$) from the overall counter circuit~($32.75\,GE$).

When injecting one fault into the netlist, \otfi finds two effective faults manipulating the output counter value without triggering the error logic.
These effective faults occur when faulting the counter increment signal or the counter clear signal, which are used by both counter instances.
Since this behavior is documented in the description of the module, we further analyze the effect of two faults into the counter.
In this setting, our tool shows in Row~\ding{172} in Table~\ref{otfi:tab:genericresults}, that $10.82\,\%$ of all injected faults are effective, \ie manipulate the output counter value to an arbitrary value but do not trigger the error signal.

\subsubsection{Double LFSR}
\label{sec:otfi:cs:prim_lfsr}
As the linear-feedback shift registers~(LSFRs) are used in \ot as the primary source of randomness, they require a strong protection against fault attacks.
The \texttt{prim\_double\_lfsr} module, which is used by several hardware IP blocks in the project, instantiates an LFSR twice and triggers an exception if the comparison of the two generated values mismatches.

In order to verify that a potentially aggressive synthesis setup does not remove the redundancy used as a fault protection, we use \otfi to induce faults into the netlist and observe the behavior of the circuit.
Here, we are interested if the error detection logic is capable of \underline{d}etecting a \underline{f}ault arbitrarily manipulating the output of the LFSR \textbf{(FD)}.
\otfi confirms that a fault into the circuit manipulating the LFSR value is detected by the error logic.
When inducing two simultaneous faults into the netlist, \otfi finds, as shown in Row~\ding{173} in Table~\ref{otfi:tab:genericresults}, $4$ effective faults either suppressing the error signal and changing the output LSFR value or manipulating the LFSR value in both LFSR modules.
Increasing the number of simultaneous faults to three increases the number of faults injected into the circuit of a size of $116.75\,GE$ to $340,692$, which takes $35\,min$ on a 16-core setup.
Based on these results, forging the LFSR output to an attacker controllable value with three or less simultaneous faults is hard to achieve.

\section{Related Work}
\label{sec:otfi:relwork}

Fault injection verification frameworks can be categorized into simulation- or verification-based approaches operating either on the RTL model or on the gate-level netlist.
As indicated in Section~\ref{sec:otfi:introduction}, frameworks~\cite{geier2020fast, jenn1995fault} working on the HDL description of a module only can provide security assumptions for this level of abstraction.
In particular, the transformation of the RTL model into the gate-level netlist, \ie the synthesis, can be responsible for inducing flaws into redundancy-based fault countermeasures by applying optimization passes.
To also detect flaws potentially introduced in this design phase, various frameworks conduct their fault experiments at the netlist level~\cite{burchard2017autofault, arribas2020cryptographic, richter2021fiver, bosio2008lifting, simevski2013automated}.
The disadvantage of simulation-based frameworks~\cite{bosio2008lifting, arribas2020cryptographic, simevski2013automated} is that they require an input stimuli covering all inputs of the circuit.
Verification-based frameworks, such as \otfi, FIVER~\cite{richter2021fiver}, and AutoFault~\cite{burchard2017autofault}, can achieve higher fault coverage as a SAT solver is responsible for probing all the undefined inputs.
Similar to \otfi, AutoFault and FIVER transform the gate-level netlist into a different representation and extract the equation of the circuit.
FIVER first transforms the circuit into a DAG and then converts this graph into a binary decision diagram to perform the symbolic fault injection.
As fault attacks originally focused on breaking cryptographic primitives, most fault injection frameworks~\cite{burchard2017autofault, arribas2020cryptographic, bosio2008lifting}, including FIVER, concentrate on analyzing such schemes.
However, when using these frameworks to analyze more generic circuits, such as a silicon design of a root-of-trust element including a broad range of fault countermeasures, there are some limitations to overcome.
First, some tools only provide support for a subset of VHDL descriptions~\cite{burchard2017autofault} and others limit the number of supported gates~\cite{richter2021fiver} to a small set.
Especially for industry-grade designs using long-established digital design flows, this constraint is severe as it is unlikely to adapt these hardware design flows.
\otfi overcomes these limitations by automatically processing and including arbitrary standard cell libraries into the framework and by translating the Verilog netlist into a unified model, \ie a directed multigraph.
This approach allows the framework to also support submodules when the boolean formula is provided.
Second, FIVER requires that the given netlist does not include any cycles, \ie the hardware designer needs to manually unroll the design before the evaluation.
As described in Section~\ref{sec:otfi:targetextraction}, \otfi is able to also handle such designs and automatically unfolds cycles found in the graph.
To overcome computational limitations for larger circuits, the architecture of \otfi makes heavy usage of multiprocessing, allowing the distribution of large workloads into the cloud.
Finally, and in contrast to other frameworks~\cite{burchard2017autofault, geier2020fast, jenn1995fault}, we release an open-source version of \otfi to encourage the verification of other security-critical designs.

Similar to related work~\cite{richter2021fiver}, \otfi can also be used to analyze the resilience of cryptographic primitives against fault attacks.
For example, when analyzing a round of the LED block cipher~\cite{guo2011led} protected by a detection-based countermeasure, the \otfi user needs to provide a plaintext-ciphertext pair in the fault configuration file.
Then, depending on the configuration, \otfi can detect \textit{(i)} whether it is possible to induce a fault with any effect on the ciphertext without triggering the countermeasure or \textit{(ii)} whether it is possible to flip certain bits in the ciphertext without triggering the countermeasure.

\section{Limitations and Future Work}
\label{sec:otfi:limit}

This section summarizes current limitations of \otfi and highlights potential future work.

\paragraph{Fault Specification.}
In the current prototype implementation of \otfi, the user needs to manually specify input and expected output values in the fault model configuration.
A possible future work could be to automate this process by parsing these values from traces generated by the simulation tools.
This parser fetches the values of the circuit of interest for a specific amount of clock cycles and automatically writes these values into a separate fault model for each clock cycle.
As \otfi is already capable of successively analyzing multiple fault models (\cf Appendix~\ref{sec:otfi:appendix_fault_model}), no additional changes in the existing framework would be required.

To assist the security engineer to specify the target circuit in the fault specification file, the \otfi repository\footnote{\link} contains an experimental feature automatically creating this file.
When using this feature, the \otfi user directly can specify the input and output nodes and their values of the target circuit in the HDL code using code annotation.
The experimental tool then extracts this annotated information from the netlist and automatically describes the target circuit in the fault specification file, \ie the tool defines the inputs and outputs and the state of the circuit. 

\paragraph{Preprocessing.}
\otfi automatically extracts a time-independent mathematical model of the circuit to analyze in the preprocessing phase.
This time-independent model is created by replacing registers used in pipeline stages with pass-through elements and by removing cycles introduced by sequential logic.
Hence, \otfi is capable of analyzing the effect of a fault in multiple clock cycles.
In addition, by automatically processing register stages and sequential loops, the framework can handle designs that the designer did not manually unroll.
However, when aiming to analyze multiple loop iterations, \eg multiple rounds in an iterative AES implementation, \otfi needs to be configured for each round individually.
Nonetheless, as \otfi allows using multiple fault configurations in a single fault model specification file executed in one verification run, this is only a minor limitation.
Nevertheless, a possible extension of the framework could automatically unroll the circuit instead of removing the loop.

\paragraph{Fault Effects \& Layout.}
\otfi and related frameworks~\cite{burchard2017autofault, arribas2020cryptographic, bosio2008lifting} model a fault at the logical and not at the electrical level.
Consequently, these frameworks cannot analyze transient faults occurring within a clock cycle and they also cannot consider the propagation delay between gates.
Additionally, these tools, including \otfi, operate on the gate-level netlist after the synthesis step and not on the layout after place and route.
As some backend tools provide the possibility to simplify and optimize the netlist before the actual place and route step, \otfi needs to be reapplied to this netlist to confirm the evaluation results.
A future work could extend \otfi to operate on the layout to also take the position of the gates into account for the analysis.

\paragraph{Performance.}
One of the main performance impact factors is the extraction and preprocessing phase (\cf Section~\ref{sec:otfi:targetextraction}).
In this phase, \otfi extracts~\textit{(i)} the target graph by finding all paths from the input and output nodes specified in the fault specification and handles~\textit{(ii)} registers used in iterative designs by finding cycles including a register.
These operations on the graph could be improved by switching to a faster Python graph library or by porting \otfi to C or C++.

Another performance limitation is the number of fault combinations, \ie fault locations and fault effects.
For an exhaustive fault analysis over all gates and multiple simultaneous injected faults, the number of fault combinations explodes.
As \otfi, for each fault combination, needs to create the differential graph, convert this graph into a boolean formula, and uses a SAT solver to evaluate the effectiveness of the faults, the number of fault combinations primarily affects the runtime.
To improve this evaluation performance, the optimizations proposed by FIVER~\cite{richter2021fiver} could be integrated.
Here, FIVER uses a fault propagation path analysis and a clustering technique to minimize the computational overhead.


\section{Conclusion}
\label{sec:otfi:conclusion}
In this paper, we presented \otfi, a pre-silicon framework capable of inducing faults and analyzing their effects on the gate-level netlist.
The framework enables hardware designers and security engineers to analyze the resilience of designs against fault attacks.
As \otfi conducts the security assessment directly on the unmodified netlist, the framework assures that \textit{(i)} the same netlist is used for the evaluation as for the next steps in the digital hardware design flow with the final tape-out step and that \textit{(ii)} potential security weaknesses still can be fixed before the chip gets manufactured.
For the evaluation, \otfi extracts the circuit of interest and injects fault into this circuit according to the fault model.
To evaluate the effect of induced faults, the framework constructs a differential graph, transforms this graph into a mathematical model, and uses a SAT solver to study the behavior of the circuit when affected by faults.
\otfi is capable of \textit{(i)} revealing whether faults affect the input-output relation of a circuit and its countermeasures and \textit{(ii)} showing whether it is possible to enter a security-critical state using a fault without triggering the countermeasures.
We utilized the framework to assess the security of several hardware modules of \ot, a secure RoT chip.
Our evaluation results presented in Section~\ref{sec:otfi:cs} showed that the unprotected AES module is highly susceptible to single faults, our proposed fault-hardening techniques increased the security, and that the other protected hardware blocks provide a strong resilience against fault attacks.

\ifanonymous
\else
\section*{Acknowledgments}
We would like to thank the anonymous reviewers and our shepherd for their valuable feedback.
Furthermore, we would like to thank Vedad Hadzic, Bettina Könighofer, and Roderick Bloem from Graz University of Technology for the initial discussion and Alphan Ulusoy from Google for the help with the implementation.
\fi

\appendix
\section{Appendix}
\subsection{Round Counter Fault Model File}
\label{sec:otfi:appendix_fault_model}
Listing~\ref{lst:otfi:faultmodelexample} shows the fault specification used for the verification of the cipher control round counter FSM discussed in Section~\ref{sec:otfi:cs:aes:ccrc}.
The \texttt{simultaneous\_faults} parameter, which also can be overwritten by a command line argument, defines the number of simultaneous faults injected into the extracted circuit.
To specify the extracted target circuit, the input and output elements need to be defined using the \texttt{stages} parameter.
These elements can be any gate, register, or input and output port of the circuit.
For the example in Listing~\ref{lst:otfi:faultmodelexample}, the circuit between the \texttt{rnd\_ctr} register with the input \texttt{Q} and the output port \texttt{D} is defined as the circuit of interest.
The \otfi tool then uses this information to automatically extract the target circuit by finding all paths between the defined input and output element.
This circuit can consist of combinational and sequential logic.
If multiple stages are provided, \otfi automatically connects them.
To constrain the SAT solver, the user needs to provide input values with the \texttt{input\_values} parameter.
In order to allow the output layer to evaluate the effect of a fault, the fault model also provides information about the expected, expected fault output value, and the alert value.
The \texttt{node\_fault\_mapping} parameter defines the mapping function of a target gate.
During the fault injection process, the boolean function of the target gate is replaced according to this mapping.
The target gate can be defined using the \texttt{fault\_locations} entry.
If the fault evaluator does not have an intuition about the critical gates which need to be analyzed, the \otfi tool is also capable of exhaustively targeting all gates in the extracted circuit.

\begin{lstlisting}[language=json, caption= {Fault specification file for the \texttt{aes\_cipher\_control\_fsm} round counter experiment.}, label={lst:otfi:faultmodelexample}]
{
    "fimodels": {
        "aes_cipher_control_fsm_rnd_cntr_target_value": {
            "simultaneous_faults": 1,
            "stages": {
                "stage_cntr": {
                    "inputs": [
                        "rnd_ctr_q_i"
                    ],
                    "outputs": [
                        "rnd_ctr_d_o"
                    ],
                    "type": "inout"
                }
            },
            "input_values": {
                "rnd_ctr_q_i": {
                    "i": {
                        "0": 1, "1": 0, "2": 0, "3": 0
                    }
                }
            },
            "output_values": {
                "rnd_ctr_d_o": {
                    "o": {
                        "0": 0, "1": 1, "2": 0, "3": 0
                    }
                }
            },
            "output_fault_values": {
                "rnd_ctr_d_o": {
                    "o": {
                        "0": 0, "1": 0, "2": 1, "3": 0
                    }
                }
            },
            "alert_values": { },
            "node_fault_mapping": {
                "NAND": [
                    "AND"
                ]
            },
            "fault_locations": {
                "Gate_189": ["stage_cntr"]
            }
        }
    }
}

\end{lstlisting}

\bibliographystyle{alpha}
\bibliography{main}

\end{document}